\documentclass{aa}

\usepackage{txfonts}
\usepackage{color, colortbl}
\usepackage[colorlinks, citecolor=blue, linkcolor=blue, urlcolor=blue]{hyperref}

\usepackage{graphicx}
\usepackage{txfonts}
\usepackage{pdflscape}  
\usepackage{color, colortbl}

\usepackage{newtxtext,newtxmath}

\usepackage[T1]{fontenc}
\usepackage{ae,aecompl}

\usepackage{pdflscape}  

\usepackage{listings}
\usepackage{pdflscape}
\usepackage{longtable}
\usepackage{wasysym}

\usepackage{graphicx}   
\usepackage{amsmath}    
\usepackage{amssymb}    
\usepackage{multirow}

\newcommand{\MS}{\ifmmode{\,}\else\thinspace\fi{\rm M}\ifmmode_{\odot}\else$_{\odot}$\fi}
\newcommand{\LS}{\ifmmode{\,}\else\thinspace\fi{\rm L}\ifmmode_{\odot}\else$_{\odot}$\fi}

\usepackage{soul}
\usepackage{blindtext}
\begin{document}

\title{The Galactic bulge exploration
\\ II. Line-of-sight velocity templates for single-mode RR~Lyrae stars}

\author{Z.~Prudil\inst{1,2}, R.~Smolec\inst{3}, A.~Kunder\inst{4}, A.~J~Koch-Hansen\inst{2}, I. D\'ek\'any\inst{2}}

\institute{
European Southern Observatory, Karl-Schwarzschild-Strasse 2, 85748 Garching bei M\"{u}nchen, Germany; \email{Zdenek.Prudil@eso.org} 
\and Astronomisches Rechen-Institut, Zentrum f{\"u}r Astronomie der Universit{\"a}t Heidelberg, M{\"o}nchhofstr. 12-14, D-69120 Heidelberg, Germany  
\and Nicolaus Copernicus Astronomical Centre, Polish Academy of Sciences, Bartycka 18, 00-716 Warszawa, Poland
\and Saint Martin's University, 5000 Abbey Way SE, Lacey, WA, 98503}

\date{\today}

\abstract
{We present a new set of tools to derive systemic velocities for single-mode RR~Lyrae stars from visual and near-infrared spectra. We derived scaling relations and line-of-sight velocity templates using both APOGEE and \textit{Gaia} spectroscopic products combined with photometric $G$-band amplitudes. We provide a means to estimate systemic velocities for the RR~Lyrae subclasses, RRab and RRc. Our analysis indicates that the scaling relation between the photometric and line-of-sight velocity amplitudes is nonlinear, with a break in a linear relation occurring around $\approx 0.4$\,mag in both the $V$-band and $G$-band amplitudes. We did not observe such a break in the relation for the first-overtone pulsators. Using stellar pulsation models, we further confirm and examine the nonlinearity in scaling relation for the RRab subclass. We observed little to no variation with stellar parameters (mass, metallicity, and luminosity) in the scaling relation between the photometric and line-of-sight velocity amplitudes for fundamental-mode pulsators. We observed an offset in the scaling relation between the observations and stellar pulsation models, mainly in the low-amplitude RR~Lyrae regime. This offset disappears when different sets of convective parameters are used. Thus, the Fourier amplitudes obtained from the photometry and line-of-sight velocity measurements can be utilized to constrain convective parameters of stellar pulsation models. The scaling relations and templates for APOGEE and \textit{Gaia} data accurately predict systemic velocities compared to literature values. In addition, our tools derived from the \textit{Gaia} spectra improve the precision of the derived systemic velocities by approximately $50$ percent and provide a better description of the uncertainty distribution in comparison with previous studies. Our newly derived tools will be used for RR~Lyrae variables observed toward the Galactic bulge.}

\titlerunning{Line-of-sight velocity templates}
\authorrunning{Prudil et al.}
\maketitle

\section{Introduction} \label{sec:Intro}

Pulsating variables of the RR~Lyrae class are horizontal branch giants associated with old stellar populations \citep{Catelan2009,Savino2020} periodically changing their brightness with periods ranging from approximately 5 to $24$ hours a day \citep{Catelan2015}. They are one of the cornerstones for studies of the structure and dynamics of the Local Group \citep[e.g.,][]{Layden1996,Clementini2001,Contreras2013,Jacyszyn-Dobrzeniecka2020,Prudil2020Disk,Savino2022}. They serve as distance indicators toward smaller stellar systems, such as globular clusters and dwarf galaxies \citep[e.g.,][]{MartinezVazquez2016,MartinezVazquez2019,Bhardwaj2021}. The RR~Lyrae stars are also used to trace tidally disrupted stellar systems often found in the Galactic halo \citep[e.g.,][]{Hendel2018,Mateu2018streams,Koposov2019Orphan,Price-Whelan2019,Prudil2021Orphan} and can help investigate the shape and mass profile of the Milky Way halo \citep[e.g.,][]{Medina2018,Erkal2019,Prudil2022,Medina2023}. Their pulsation properties permit estimating distances with a precision better than 5 percent \citep[e.g.,][]{Catelan2004,Braga2015,Neeley2017,Muraveva2018Ret} and deriving photometric metallicity estimates for a population of RR~Lyrae stars \citep[e.g.,][]{Jurcsik1996,Smolec2005,Dekany2021}.

The obvious advantages of utilizing the pulsation properties of RR~Lyrae stars come at the cost of the intricacies of deriving information from stellar spectra compared to nonvariable stars. A given RR~Lyrae star (depending on its pulsation amplitude) undergoes changes in its temperature ($\sim 1000$\,K), and surface gravity ($\sim 1$\,dex) \citep[see, e.g.,][]{Cacciari1992,Skillen1993,For2011chem,Pancino2015}. In addition, its observed line-of-sight velocity varies as well ($\sim 60$\,km\,s$^{-1}$, depending on the spectral lines used for estimating the line-of-sight velocity) within its single pulsation cycle \citep[e.g.,][]{Clementini1990,Jeffery2007,Sesar2012,Braga2021}. This complicates the co-addition of nonconsecutively observed spectra, and the rapid changes in the line-of-sight velocity put constraints on the exposure time (longer exposures would result in asymmetric lines). Particularly, the variation in line-of-sight velocity where the pulsation of the atmosphere is convolved with the center-of-mass motion (from here on referred to as the "systemic velocity") needs to be taken into account before one uses measured line-of-sight velocities of RR~Lyrae for kinematical studies. 

There are several approaches to estimating the systemic velocity of a given RR~Lyrae star. First, one can sample the entire line-of-sight velocity variation and calculate the mean velocity. This method is costly with regard to the observation time and may not be possible for some RR~Lyrae variables with long pulsation periods or periods of $0.5$\,days, especially with ground-based observations. Second, assuming that the pulsation ephemerides such as pulsation period and reference time are known, one can time the observing window for a phase around 0.4 - 0.5 where the line-of-sight velocity is close to the actual systemic velocity \citep{Liu1991,Sesar2012}. Although less costly in observation time, this method requires preparation before observation and the capability to modify the observation plan, which is often not possible for extensive spectroscopic surveys \citep{4MOST2014,WEAVE2014,Zasowski2017,Blanton2017,Kollmeier2017}. The third option is to observe the star at any point in its pulsation cycle and then later shift the observation along a velocity template to obtain its systemic velocity. This method, similar to the second one, requires prior knowledge of the pulsation ephemerides and the pulsation amplitude as well \citep[from photometric observations, often from the $V$ passband; e.g.,][]{Liu1991,Sesar2012,Braga2021}. Unlike the previous two methods, it does not constrain the spectroscopic observations. One can determine a systemic velocity from a single line-of-sight measurement at any point during the pulsation cycle. 

The third approach seems to be the most used in the past decade \citep[e.g.,][]{Sesar2013Orphan,Kunder2016,Kunder2020,Liu2020,Hanke2020,Prudil2021Orphan,Medina2023}. In addition to the pulsation ephemerides, the last method relies on the scaling relation between amplitudes of photometric and line-of-sight velocity curves and a line-of-sight velocity template that models the behavior of line-of-sight velocities during the pulsation cycle. From the previous studies, scaling relations and templates are available for the most prominent spectral hydrogen lines (Balmer lines H$\alpha$, H$\beta$, H$\gamma$, H$\delta$) and metallic lines (e.g., Fe, Mg, etc.). In this work, we focus on expanding the scaling relations and line-of-sight velocity templates toward redder wavelengths, particularly toward the calcium triplet and $H$-band covered by the surveys described below. We selected these two regions since we aim to use newly derived scaling relations and templates to study the Galactic bulge, as most of the spectroscopic data for RR~Lyrae stars cover this approximate redder wavelength range.

One of the two spectroscopic surveys used in this work is the Apache Point Observatory Galactic Evolution Experiment \citep[APOGEE;][]{Majewski2017,Wilson2019APO,Beaton2021}. APOGEE is a large-scale infrared spectroscopic survey of the Milky Way focusing mainly on the highly reddened regions, such as the Galactic disk and Galactic bulge. The observations were conducted at the Apache Point Observatory (APO) in New Mexico, USA, and at the Las Campanas Observatory (LCO) in the Atacama Desert in Chile. The APOGEE survey covers the $H$ passband ($1.51$--$1.70$\,$\mu$m wavelength range) with a resolution of $\sim22500$ \citep[SDSS;][]{Eisenstein2011SegueII,Blanton2017}. The main targets of the APOGEE survey are giant stars, for which APOGEE derived stellar atmospheric parameters, line-of-sight velocities, and chemical abundances for numerous elements\citep{Nidever2015,Abdurrouf2021APOGEEDR17}. Among their targets toward the inner Galaxy, over $6000$ RR~Lyrae stars were observed, some of which have several epochs of spectroscopic measurements. Therefore, obtaining their systemic velocities could aid in studies of the Galactic bulge formation history.

The second survey we used is the \textit{Gaia} space telescope mission focused on measuring objects' precise positions and motions \citep{Gaia2016,GaiaDR32023}. One of the instruments on board \textit{Gaia} is the Radial Velocity Spectrometer (RVS), which aims to provide multi-epoch line-of-sight velocities. The RVS resolving power is $R \approx 11500$, with a wavelength coverage ranging from $8450$\,\AA~to $8720$\,\AA,~ thus covering the calcium triplet (Ca\,T) region \citep{Cropper2018,Katz2022}. The latest data release of the \textit{Gaia} catalog provides individual line-of-sight velocities for $1096$ RR~Lyrae stars, with $1086$ having more than seven observations \citep{Clementini2023}. The uncertainties on individual line-of-sight velocities range from $\approx 0.5$\,km\,s$^{-1}$ up to $\approx 85$\,km\,s$^{-1}$ \citep{Clementini2023}. This catalog provides exquisite data for calibrating scaling relations between photometric and line-of-sight velocity amplitudes and creating line-of-sight velocity templates. The tools mentioned above can be used for estimating systemic velocities for RR~Lyrae stars observed by the Bulge RR~Lyrae Radial Velocity Assay \citep[BRAVA-RR,][]{Kunder2016,Kunder2020}.

Unlike previous studies that provided linear (with both a slope and intercept) scaling relations between photometric and line-of-sight velocity amplitudes \citep[e.g.,][]{Jones1988,Liu1991,Sesar2012,Braga2021}, in our work the intercept of the scaling relations was set to zero. This conserves the physical representation of the relations and avoids deriving transformation equations that would predict line-of-sight velocity changes with zero photometric amplitude variation. Section~\ref{sec:InfraredSpec} describes the derivation of the line-of-sight velocity templates and the scaling relations from the APOGEE spectra. Section~\ref{sec:GaiaCaT} focuses on a similar task but for the \textit{Gaia} RVS spectra that cover the Ca\,T region. In Section~\ref{sec:NonLinearity}, we discuss nonlinearity in scaling relations for fundamental-mode RR~Lyrae stars, and we compare our observational data with stellar pulsation models. In Section~\ref{sec:ExampleOfAnalysisAndTesting}, we outline an example of the determination of a systemic velocity using our derived templates and scaling relations together with their testing. Last, Section~\ref{sec:Summary} provides a summary of our results.


\section{Line-of-sight velocity templates for APOGEE spectra} \label{sec:InfraredSpec}

This section describes our approach to deriving the amplitude scaling relations and line-of-sight velocity templates for RR~Lyrae stars. We focus on single-mode RR~Lyrae stars -- those pulsating in the fundamental mode, RRab, as well as those pulsating in the first overtone, RRc -- by using APOGEE near-infrared spectra for RR~Lyrae stars in the Solar neighborhood. In deriving the aforementioned tools, we have drawn inspiration from the studies by \citet{Sesar2012} and \citet{Braga2021} that undertook a similar process but in the visual part of the spectra. 

APOGEE provides various spectroscopic products. In particular, we were interested in the individual visits (\texttt{apVisit}), which are a combination of approximately eight individual observations (observed back-to-back), each with $500$\,s exposures. These individual observations comprise the foundation for creating the line-of-sight velocity templates and constructing the amplitude scaling relations for RR~Lyrae pulsators observed by the APOGEE survey. 

\subsection{Obtaining the line-of-sight velocities} \label{subsec:TempAPO}

The APOGEE survey intends to observe approximately $10000$ RR~Lyrae stars both on the southern and northern sky \citep{Bowen1973,Gunn2006,Holtzman2010,Zasowski2017}. In the northern hemisphere, a small sample of $30$ nearby ($H < 10$\,mag) RR~Lyrae variables were observed by APOGEE to obtain, at least for some of them, sufficient coverage of the entire pulsation cycle that would then serve in creating the line-of-sight velocity template (see Table~\ref{tab:30RR} for their full list). These RR~Lyrae stars have been marked with the \texttt{rrlyr} label in the APOGEE data products. The APOGEE survey serendipitously also observed other nearby RR~Lyrae stars \citep[listed as RR~Lyrae stars in the International Variable Star Index, VSX, database\footnote{\url{https://www.aavso.org/vsx/index.php}},][]{Watson2006VSX}. We decided to utilize some of these RR~Lyrae stars (we used only those with at least five visits and S/N $>50$) to create the line-of-sight velocity templates for both RR~Lyrae subclasses and the relation between photometric amplitudes and line-of-sight velocity amplitudes (a list of these RR~Lyrae stars used in the creation of the line-of-sight-velocity templates is given in Table~\ref{tab:OtherRR}).

As an initial step, we acquired the individual exposures from the SDSS ftp data access.\footnote{\url{https://dr17.sdss.org/sas/dr17/apogee/spectro/redux/dr17/visit/}} The APOGEE spectra are provided in vacuum wavelengths \citep{Majewski2017}; therefore, we transformed them into the air wavelengths using a relation from the SDSS website.\footnote{\url{https://www.sdss.org/dr17/irspec/spectra/}} For each exposure, we determined the barycentric observation time $T_{\rm BJD}$. To determine reliable barycentric line-of-sight velocities ($v_{\rm los}$), we used the \texttt{iSpec} package \citet{Blanco2014,Blanco2019iSpec} to synthesize a grid of synthetic spectra that cover common physical properties of RR~Lyrae stars \citep[based on stellar parameters found in the literature,][]{For2011chem,Sneden2017,Preston2019}. We used the following set of physical properties to create synthetic spectra: {[Fe/H]$ = (-2.5, -2.0, -1.5, -1.0, -0.5, 0.0)$\,dex};
$T_{\rm eff} = (6000., 6500., 7000., 7500.)$\,K; log\,$g = (2.0, 2.5, 3.0)$\,dex; and a microturbulence velocity of $\xi_{\rm turb} = 3.5$\,km\,s$^{-1}$ \hspace{1cm}.

The spectra were synthesized using the radiative transfer code MOOG \cite[February 2017 version,][]{Sneden1973} implemented in \texttt{iSpec}, the ATLAS9 model atmosphere \citep{Castelli2003}, the solar reference scale from \citet{Asplund2009}, and the line list from the VALD database.\footnote{\url{http://vald.astro.uu.se/}} The spectral range was tailored to APOGEE spectra (from $15000$\,\AA~to $17000$\,\AA). Thus, we covered the entire APOGEE spectral range and did not focus on small spectral features, such as specific spectral lines \citep[e.g., studies by][]{Sesar2012,Braga2021}. This decision was motivated by the intensities of the hydrogen lines, members of the Paschen and Brackett series that dominate this region of spectra in RR~Lyrae stars. These hydrogen lines are the only ones seen in low signal-to-noise APOGEE spectra of RR~Lyrae variables toward the Galactic bulge.



To determine the line-of-sight velocity ($v_{\rm los}$), we cross-correlated each spectrum with synthesized templates, selecting the one that minimized the error in $v_{\rm los}$. This template was then used in a Monte Carlo simulation ($200$ iterations for each measurement) to address the uncertainties in the observed spectra. We adjusted flux values according to their normally distributed errors. Finally, the bootstrap method was applied to this data to calculate the average and standard deviation of $v_{\rm los}$ and its variation, $\sigma_{v_{\rm los}}$. To remove erroneous measurements, we applied a condition on the $v_{\rm los}$ and on its uncertainty:
\begin{equation} \label{eq:CritTempAPO}
\left| v_{\rm los} / \sigma_{v_{\rm los}} \right| > 5 \hspace{1cm} \text{or} \hspace{1cm} \sigma_{v_{\rm los}} < 10~\mathrm{km\,s^{-1}}.
\end{equation}

\subsection{Estimating the line-of-sight velocity amplitudes} \label{subsec:TempAPOAMP}

Once we acquired $v_{\rm los}$ and $T_{\rm HJD}$ for the nearby RR~Lyrae stars, we focused on obtaining photometric information to constrain the basic pulsation properties of our RR~Lyrae template dataset. We used photometric data from the \textit{Gaia} RR~Lyrae catalog \citep{Clementini2023}, which provides photometry in $G$, $G_{\rm RP}$, and $G_{\rm BP}$ as well as pulsation properties (pulsation period $P$, times of brightness maxima $E$, and amplitudes of the light and velocity changes Amp$_{G}$, Amp$_{G_{\rm RP}}$, Amp$_{G_{\rm BP}}$, and Amp$_{RV}$).

Once we obtained the \textit{Gaia} ephemerides ($P$ and $E$) for the stars in our dataset, we phased the observed $v_{\rm los}$ using the element-wise remainder of division of the product, and we assessed the quality of the line-of-sight velocity curves for our RR~Lyrae template dataset. We selected only those stars for which both the light descending side and light ascending side of the line-of-sight velocity curve were covered based on visual inspection of the data. In the end, we had $13$ RR~Lyrae stars (eight RRab and five RRc stars) that were suitable for estimating line-of-sight velocity amplitudes (Amp$_{\rm los}$) and subsequently creating the line-of-sight velocity templates. We refer to these stars as our APOGEE template dataset, and they are marked with an asterisk in Tables~\ref{tab:30RR} and \ref{tab:OtherRR}. Our dataset of $13$ template stars is somewhat similar in number to that of \citet{Sesar2012}, who used six RRab stars to construct radial velocity templates, and \citet{Braga2021}, who used $31$ RRab but five RRc stars to obtain radial velocity templates.

In the next step, we estimated the line-of-sight velocity amplitudes. Despite using individual exposures, our line-of-sight velocity curves do not cover the entire phase curve without small gaps, and the curves can have a low number of points (e.g., $27$ reliable measurements for the prototype of the RR~Lyrae variable class). Therefore, a direct Fourier fitting \citep[as used in ][]{Braga2021} could lead to problems in obtaining a reliable Amp$_{\rm los}$. Thus, similar to \citet{Dekany2021}, we turned to the Gaussian process regression \citep[GPR;][]{Rasmussen2005} as a means to describe phased curves of RR~Lyrae stars (but in our case, we work with the line-of-sight velocity curves). We used phase-folded line-of-sight velocities in conjunction with the Gaussian process (GP) regressor implemented in the \texttt{george} module \citep{Ambikasaran2015george}.\footnote{Available here: \url{https://george.readthedocs.io/en/latest/}.} For the GPR, we used the following kernels:
\begin{lstlisting}[basicstyle=\footnotesize, language=python, mathescape=true]
kernel = ExpSquaredKernel(hyperparameters) $\times$
          Matern32Kernel(hyperparameters). $\hspace{1cm}$
\end{lstlisting}
The hyperparameters for each kernel were optimized through the marginalized log-likelihood function.\footnote{Using the procedure described here: \url{https://george.readthedocs.io/en/latest/tutorials/hyper/}.} 

\begin{figure}
\includegraphics[width=\columnwidth]{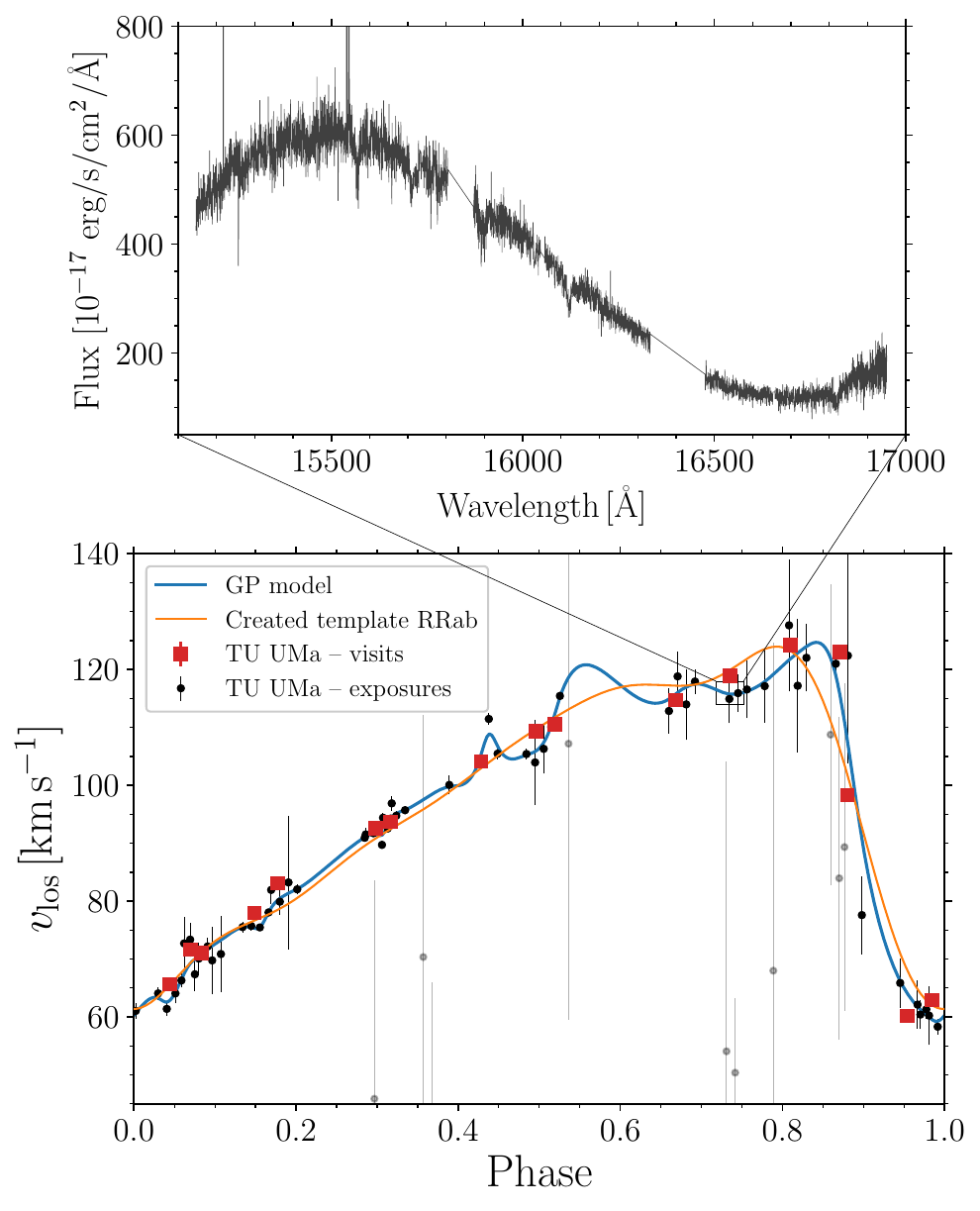}
\caption{Example of a single exposure spectrum (top panel) and line-of-sight velocities (bottom panel) used in this work. The top panel depicts a spectrum of TU-UMa taken at $\text{HJD}=2457484.880667$\,day. The bottom panel shows the line-of-sight velocities derived in this work (black dots) and the velocities extracted from APOGEE data products (red squares). The gray dots represent estimated velocities that did not fulfill the criteria in Eq.~\ref{eq:CritTempAPO}.}
\label{fig:ExampleOfAPORAD}
\end{figure}

In Figure~\ref{fig:ExampleOfAPORAD}, we depict a phased line-of-sight velocity curve based on the determined $v_{\rm los}$ from individual exposures, line-of-sight velocities from the co-added visits, and a fit of the GP regressor for estimating the amplitude of the line-of-sight velocity changes. The GP model for each individual star does introduce some curves along the rising branch, likely due to the gaps in the observations along the full pulsation cycle, but much of this is smoothed out when the full sample of stars is combined in the generation of the final RRab and RRc template. In Table~\ref{tab:PulsPropTEMPAPO}, we list the basic pulsation properties and the line-of-sight velocity amplitudes\footnote{We calculated the velocity amplitudes from the GPR models as a difference between the maximum and minimum value of the line-of-sight velocity of a given pulsator.} for stars used in the creation of the amplitude scaling relation and the $v_{\rm los}$ curve template.

\setlength{\tabcolsep}{4.75pt}
\begin{table}
\caption{List of nearby RR~Lyrae variables used in the creation of the line-of-sight velocity template and the scaling relation between Amp$_{\rm los}$ and Amp$_{G}$. The first column presents the names of the stars. Columns two and three contain the RR~Lyrae subclass and the line-of-sight velocity amplitude. The last two columns contain the equatorial coordinates of the sample stars.}
\label{tab:PulsPropTEMPAPO}
\begin{tabular}{lrlrr}
\hline
Alt. ID & Type & Amp$_{\rm los}$  & R.A.        & DEC  \\ \hline
SW-And & RRab & $59.5 \pm 1.0$        & 5.92954   & 29.40100  \\
YZ-Hyi  & RRab & $57.6 \pm 3.4$        & 6.09269   & -77.36900 \\
CO-Tuc  & RRc  & $26.1 \pm 1.7$        & 7.14105   & -72.16911 \\
TT-Lyn & RRab & $51.2 \pm 3.2$        & 135.78200 & 44.58560  \\
T-Sex  & RRc  & $23.6 \pm 1.4$        & 148.36800 & 2.05732   \\
TU-UMa & RRab & $65.7 \pm 6.8$        & 172.45200 & 30.06730  \\
CN-Cam & RRab & $32.1 \pm 3.0$        & 174.04900 & 81.29360  \\
UU-Vir  & RRab & $64.6 \pm 3.3$        & 182.14614 & -0.45676  \\
UV-Vir  & RRab & $59.7 \pm 4.9$       & 185.31972 & 0.36750   \\
OW-Dra  & RRc  & $25.4 \pm 2.1$        & 195.87959 & 71.11225  \\
RR-Lyr& RRab & $64.9 \pm 1.2$       & 291.36600 & 42.78440  \\
DH-Peg & RRc  & $27.2 \pm 1.2$        & 333.85699 & 6.82263   \\
RZ-Cep & RRc  & $28.8 \pm 1.9$        & 339.80499 & 64.85850  \\ \hline
\end{tabular}
\end{table} %

\subsection{Scaling relation between Amp$_{\rm los}$ and Amp$_{G}$ based on APOGEE spectra} \label{subsec:TempAPOAMPscal}

Using the photometric amplitudes obtained from the \textit{Gaia} catalog, Amp$_{G}$, and the acquired Amp$_{\rm los}$ from the APOGEE, we derived the scaling relation between both parameters. Previous studies, such as \citet{Sesar2012} and \citet{Braga2021}, used a linear scaling relation between the photometric and velocity products. We derived separate amplitude relations for both RR~Lyrae subclasses. In the case of RRc-type stars, we also used a linear relation, but for RRab, we decided to use a second-degree polynomial relation. This is supported by the observed trend for RRab stars in APOGEE data and subsequently confirmed in Section~\ref{sec:GaiaCaT} for \textit{Gaia} RVS spectra.

We implemented a fitting procedure aimed at optimizing the following relations:
\begin{gather} \label{Eq:ApogeeFit}
\text{Amp}_{\rm los}^{\rm RRab} = x_{0} \cdot \text{Amp}_{G}^{2} + x_{1} \cdot \text{Amp}_{G} \\
\text{Amp}_{\rm los}^{\rm RRc} = x_{0} \cdot \text{Amp}_{G} \hspace{1cm},
\end{gather}
where $x_{0}$ and $x_{1}$ are parameters of the fit (we omitted the intercept to preserve the physical meaning of the relation). In the search for optimal parameters and their covariances, we utilized the \texttt{emcee} module \citep{Foreman-Mackey2013}. For each RR~Lyrae subclass, we ran the Markov chain Monte Carlo simulation with $200$ walkers for $30000$ samples. For deriving the best-fit values, we thinned the sample by $\tau=20$ and marked the initial $10000$ samples as burn-in. 

\begin{figure*}
\includegraphics[width=2\columnwidth]{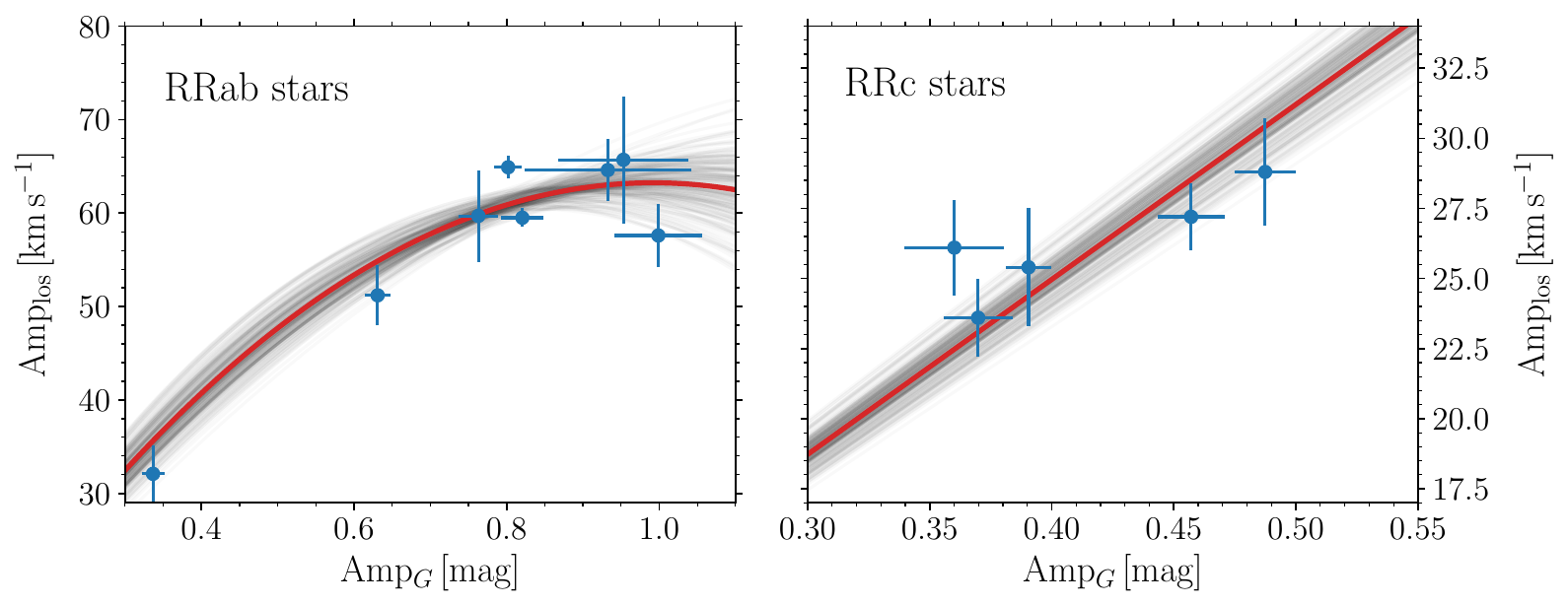}
\caption{Scaling relations between the photometric and line-of-sight velocity amplitudes for the fundamental-mode RR~Lyrae stars (left-hand panel) and first-overtone RR~Lyrae variables (right-hand plot). Both RR~Lyrae subclasses are depicted using blue points. The best-fit parameters are shown with a solid red line, while gray lines represent their variations.}
\label{fig:ScallingRelaFIG}
\end{figure*}

Figure~\ref{fig:ScallingRelaFIG} shows the scaling relations for both single-mode RR~Lyrae stars. The scaling relations for the RRab and RRc pulsators, together with their uncertainties, can be expressed with the following equations:
\begin{gather} \label{eq:ScalRel}
\text{For RRab stars:}\hspace{0.25cm}\text{Amp}_{\rm los} = -64 \cdot \text{Amp}_{G}^{2} + 127 \cdot \text{Amp}_{G}\\
\text{For RRc\phantom{b} stars:}\hspace{0.25cm}\text{Amp}_{\rm los} = 62(2) \cdot \text{Amp}_{G} \hspace{1cm}.
\end{gather}
For RRab variables, we also include a covariance matrix for the uncertainty estimation:
\begin{equation}
\begin{split}
\text{Cov}^{\rm RRab} = 
\begin{bmatrix} 
120 & -98 \\
-98 & 81 \\
\end{bmatrix}
\end{split} \\.
\end{equation}

\subsection{Line-of-sight velocity templates for APOGEE spectra} \label{subsec:TempAPOAA}

When creating the line-of-sight velocity templates for RRab and RRc stars, one needs first to normalize the line-of-sight velocity curves of the stars. We utilized the phased $v_{\rm los}$ curves and their associated GPR models. We used GPR models for the individual calibration stars to normalize the phase-folded $v_{\rm los}$ curves (between $-0.5$ and $0.5$). The individual normalized phased curves were then co-added into two normalized curves, one for each RR~Lyrae subclass. Unlike previous studies, we determined the systemic velocity directly from the shift of the line-of-sight velocity template and not from a specific phase point \citep[e.g., $0.38$ or $0.5$; see][]{Liu1991,Sesar2012}.

\begin{figure*}
\includegraphics[width=2\columnwidth]{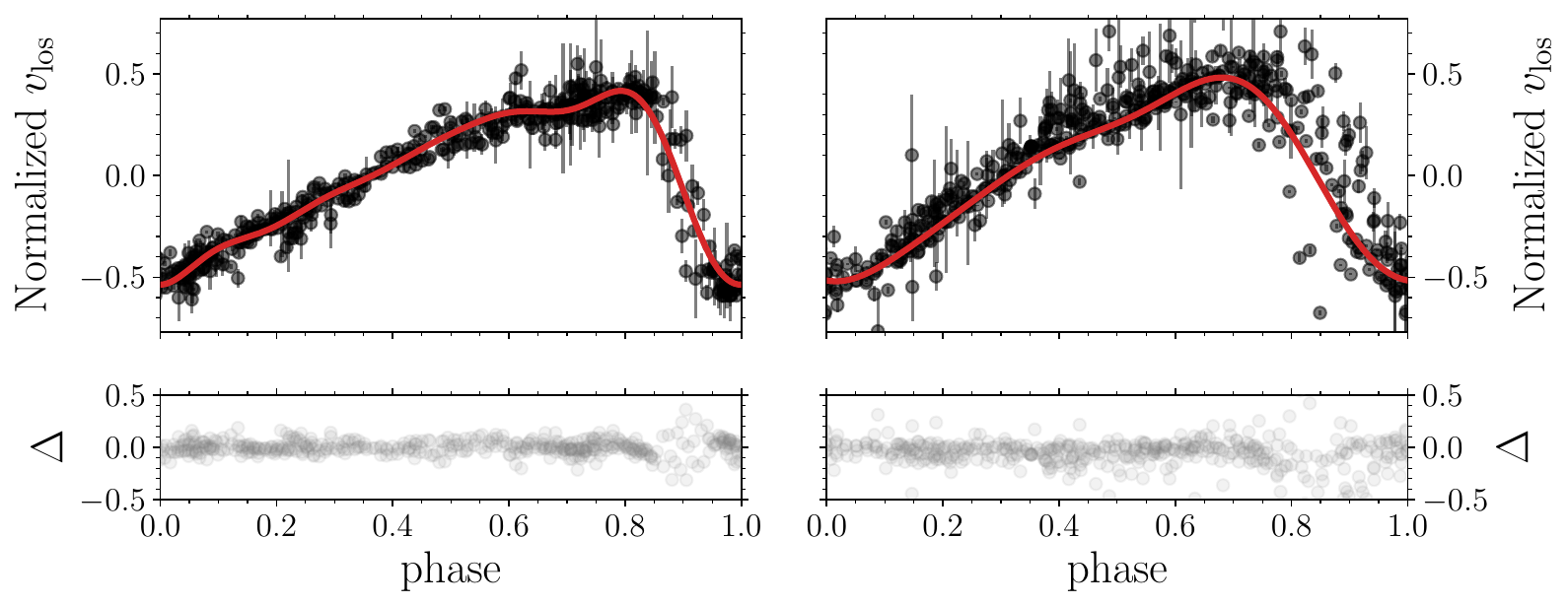}
\caption{Line-of-sight velocity templates for the RRab (left-hand panel) and RRc (right-hand plots) pulsators. The black points in the top panels represent individual measurements of $v_{\rm los}$, and the solid red lines depict the Fourier model for each normalized line-of-sight velocity curve. The bottom panels show the difference, $\Delta$, between the model (Fourier fit) and the phased $v_{\rm los}$ data for both RR~Lyrae subclasses.}
\label{fig:tempAPO}
\end{figure*}

These template curves, as we refer to them, were then modeled using Fourier decomposition with the following relation:
\begin{equation} \label{eq:FourSerTemp}
f\left ( \varphi \right ) = a_{0} + \sum_{k=1}^{n} a_{k} \cdot \text{sin} \left [2\pi k \phi \right ] + a_{k+1} \cdot \text{cos} \left [2\pi \left( k+1 \right) \phi \right ] \\,
\end{equation}
where $\phi$ represents the time element (in this case, phase) and $a_k$ represents the amplitudes of the individual Fourier components. For RRab and RRc stars, we selected the fifth ($n=5$)  and third degree ($n=3$) of the Fourier series, respectively. In Figure~\ref{fig:tempAPO}, we depict the phase-folded normalized line-of-sight velocity curves with associated Fourier templates. In Table~\ref{tab:FourTEMPApo}, we list the Fourier coefficients for both models of RR~Lyrae subclasses. The $a_{0}$ coefficients are zero due to the prior normalization of the line-of-sight velocity curves.

In the bottom panels of Fig.~\ref{fig:tempAPO}, we observed a noticeable scatter in the template curves, particularly between phases of rapid contraction ($0.8$ to $1.0$). We included this scatter in $v_{\rm sys}$ determination (see example in Section~\ref{subsec:ExampleOfAnalysis}) as a root-mean-square scatter in the template, which is similar to the approach described in \citet{Sesar2012}. This can be represented as a spline function describing the scatter along the pulsation phase (see Table~\ref{tab:TEMPerr}). 

\begin{table*}
\caption{List of Fourier coefficients for RR~Lyrae line-of-sight velocity templates for APOGEE spectra.}
\label{tab:FourTEMPApo}
\begin{tabular}{lccccccccccc}
\hline
Template  & $a_0$     & $a_1$      & $a_2$      & $a_3$      & $a_4$      & $a_5$     & $a_6$      & $a_7$     & $a_8$      & $a_9$     & $a_{10}$    \\ \hline
$\texttt{Temp}^{\rm RRab}$ & $0.0000$ & $-0.2667$ & $-0.2806$ & $-0.0351$ & $-0.1358$ & $0.0222$ & $-0.0884$ & $0.0455$ & $-0.0300$ & $0.0192$ & $-0.0016$ \\
$\texttt{Temp}^{\rm RRc}$  & $0.0000$ & $-0.2278$ & $-0.3871$ & $0.0265$ & $-0.1294$ & $0.0302$ & $0.0006$ & --      & --       & --      & -- \\ \hline
\end{tabular}
\end{table*}

\section{Gaia calcium triplet spectra} \label{sec:GaiaCaT}

The \textit{Gaia} RR~Lyrae catalog provides all the necessary data to develop scaling relations and line-of-sight velocity templates for the Ca\,T region. The abundance of data allowed for the creation of a tailored method for Ca\,T spectra observed by the BRAVA-RR survey for RR~Lyrae pulsators toward the Galactic bulge \citep{Kunder2016,Kunder2020}. 

Using photometric amplitudes from the \textit{Gaia} survey for calibration of the scaling relations comes with a trade-off. \textit{Gaia} provides exceptional photometry for hundreds of thousands of RR~Lyrae stars taken simultaneously with the \textit{Gaia} RVS spectra. However, approximately $20$ percent of the RR~Lyrae stars in the Galactic bulge \citep{Clementini2023}, which are covered by the APOGEE and BRAVA-RR surveys, do not have photometric amplitudes in \textit{Gaia} passbands. The main source of photometric amplitudes for these stars comes from the fourth data release of the Optical Gravitational Lensing Experiment \citep[OGLE-IV;][]{Soszynski2014BulgeRRlyr,Udalski2015,Soszynski2019Disk}. OGLE-IV provides a detailed sample of RR~Lyrae stars in the Galactic bulge, including abundant photometry in the $V$ and $I$ passbands.

Thus, we decided to create conversion relations between the $V$, $I$, and $G$ passbands to enable transitions to and from the \textit{Gaia} $G$-band in cases where amplitudes in other passbands are available. To this end, we crossmatched the \textit{Gaia} RR~Lyrae catalog with the OGLE-IV catalog using equatorial coordinates (within a $1$ arcsecond radius). This crossmatch provided us with photometry for Galactic bulge RR Lyrae stars in the $V$ and $I$ passbands. We utilized our fitting routine (similar to the one used in Eq.~\ref{eq:FourSerTemp}) to decompose the OGLE-IV light curves using the Fourier series. The following Fourier light curve decomposition was optimized: 
\begin{equation} \label{eq:FourSer}
m\left ( t \right ) = m_{V} + \sum_{k=1}^{n} A_{k} \cdot \text{cos} \left (2\pi k \vartheta + \varphi_{k} \right ) \\.
\end{equation}
In equation~\ref{eq:FourSer}, $m_{V}$ represents the mean apparent magnitude, and $A_{k}$ and $\varphi_{k}$ stand for amplitudes and phases, respectively. The term $n$ denotes the degree of the fit that we adapted for each light curve using the same approach as in \citet{Prudil2019OOspat}. Here, $\vartheta$ represents the phase function, which is defined as:
\begin{equation} \label{eq:phasing}
\vartheta = \left(\text{HJD}-E\right)/P \\,
\end{equation}
where HJD is the Heliocentric Julian Date of the observation and $E$ stands for the time of brightness maximum. The time of brightness maxima was used as $E$ to set a common light curve reference point.

{To create conversion relations, we followed the same procedure as in Subsection~\ref{subsec:TempAPOAMPscal} for the first-overtone pulsators, selecting only a linear fit without an intercept. To exclude potentially blended stars, we applied the condition \texttt{ipd\_frac\_multi\_peak}~$ < 5$. We fitted amplitude comparisons for RRab- and RRc-type stars separately, and the derived relations are listed below:
\begin{gather} \label{eq:AmpsConversion}
\text{RRab:}\hspace{0.2cm}\text{Amp}_I = 0.80(0.01) \cdot \text{Amp}_G \\ \label{eq:AmpsConversionVG}
\text{RRab:}\hspace{0.2cm}\text{Amp}_V = 1.32(0.01) \cdot \text{Amp}_G \\
\text{RRc:}\hspace{0.2cm}\text{Amp}_I = 0.74(0.02) \cdot \text{Amp}_G \label{eq:AmpsConversionVGRRc}\\
\text{RRc:}\hspace{0.2cm}\text{Amp}_V = 1.33(0.04) \cdot \text{Amp}_G \label{eq:AmpsConversionVGRRc} \hspace{1cm}. 
\end{gather}
The derived conversion relations were also used in the cases where we needed to compare scaling relations optimized on $V$ passband.

\subsection{Scaling relation between $\text{Amp}_{G}$ and $\text{Amp}_{\rm los}$ derived from \textit{Gaia} spectra} \label{subsec:ScalRelGaia}

Our method for obtaining scaling relations and line-of-sight velocity templates closely mirrors the process we utilized for the APOGEE dataset. We extracted pulsation properties from the \textit{Gaia} catalog (pulsation period and epoch of the maximum of the light curve in the $G$-band) for all single-mode RR~Lyrae stars with publicly available individual $v_{\rm los}$ and their associated uncertainties (in total, $1086$ variables). We implemented the same criteria for the line-of-sight velocities as for APOGEE (see Eq.~\ref{eq:CritTempAPO}). Moreover, we selected RR~Lyrae stars that had at least $20$ \textit{Gaia} $v_{\rm los}$ measurements (after applying criteria on the line-of-sight velocities), and we only selected RR~Lyrae stars for which the measured Amp$_{\rm los}$ fulfilled the following criteria: 
\begin{equation} 
\left| \text{Amp}_{\rm los} / \sigma_{\text{Amp}_{\rm los}} \right| > 4 \hspace{0.7cm} \text{or} \hspace{0.7cm} \sigma_{\text{Amp}_{\rm los}} < 10~\mathrm{km\,s^{-1}} \\.
\end{equation}
Our last criterion was to only use stars above \textit{Gaia} flag \texttt{ipd\_frac\_multi\_peak}~$ < 5$. The last criterion was followed in order to reduce the possible blending of stars, as it would decrease pulsation and velocity amplitudes. In total, for RRab- and RRc-type stars, we had $245$ and $88$ $\text{Amp}_{\rm los}$ to derive scaling relations, respectively.

As in the APOGEE case, we used a GP regressor to calculate $\text{Amp}_{\rm los}$. While modeling the amplitude scaling relation, we noticed a non-negligible scatter in the data, and we included intrinsic scatter in the fit (for both RRab and RRc variables, $\varepsilon_{\text{RRab}}$ and $\varepsilon_{\text{RRc}}$). To further explore the nature of the scatter, we crossmatched our fundamental mode dataset with a catalog of known RRab stars exhibiting the Blazhko effect \citep{Skarka2013}. We found that approximately 9 percent of our RRab sample shows signs of modulation, and thus, some of the scatter in the RRab scaling relation may be due to the Blazhko effect. Including or removing the identified Blazhko stars did not have an effect on the scatter in the scaling relation to RRab stars. This was partially caused by the fact that we do not have complete information on the identification of Blazhko stars in the Solar neighborhood.

In Figure~\ref{fig:ScallingRelGaia}, we depict the scaling relation for both RR~Lyrae subclasses. It is typically assumed that there is a linear relation between $\text{Amp}_{G}$ and $\text{Amp}_{\rm los}$ (for both RR~Lyrae subtypes), but it is apparent that the amplitudes do not follow a linear relation (similar to the case we saw in Section~\ref{subsec:TempAPOAMPscal}). Instead, there is a break at $\text{Amp}_{G} \approx 0.4$\,mag, where stars with an $\text{Amp}_{G}$ smaller than $0.4$ have a steeper scaling relation than those with $\text{Amp}_{G} > 0.4$\,mag. For comparison, we overplotted \textit{Gaia} data and our derived relation for RRab stars with linear relations found in the literature \citep[][Fe lines;]{Liu1991,Sesar2012,Braga2021}. For interested readers, in the Appendix, we enclosed a Table~\ref{tab:SevenLow} with $34$ RRab variables that are located at the nonlinear part of the scaling relation with some of their basic properties. We observed that for an RR~Lyrae variable with $\text{Amp}_{G} = 1.0$\,mag, the scaling relations from the literature \citep[based on Fe lines and converted from Amp$_{V}$ to Amp$_{G}$;][]{Liu1991,Sesar2012,Braga2021} together with our relation predict $\text{Amp}_{\rm los} \approx 68$\,km\,s$^{-1}$. On the opposite side of the amplitude distribution, for low amplitudes (e.g., $\text{Amp}_{G} = 0.25$\,mag), our relation predicts amplitudes lower than $\text{Amp}_{\rm los} \approx 10$\,km\,s$^{-1}$ 
(for \citeauthor{Liu1991}~\citeyear{Liu1991}, \citeauthor{Sesar2012}~\citeyear{Sesar2012} and \citeauthor{Braga2021}~\citeyear{Braga2021}). One consequence of a steeper relation between $\text{Amp}_{G}$ and $\text{Amp}_{\rm los}$ at a smaller $\text{Amp}_{G}$ is that the intercept of the scaling relation naturally approaches zero, which is what is physically expected -- a photometric amplitude of zero should also correspond to no $\text{Amp}_{\rm los}$. On the other hand, for the first-overtone pulsators, we did not see a rapid decline of $\text{Amp}_{\rm los}$ toward smaller photometric amplitudes. The range of both $G$-amplitudes and $\text{Amp}_{\rm los}$ is considerably smaller than for the RRab pulsators, which may be a factor contributing to more constant scaling relations. Also, the scaling relation of RRc pulsators  naturally passes through an intercept of zero without the need for a change in slope. More RRc-type variables at the low-amplitude end would help solidify the use of a linear relation instead of a higher degree polynomial. 

The scaling relations between $\text{Amp}_{\rm los}$ and $\text{Amp}_{G}$, their associated covariance matrix, and uncertainty are described in the following equations:
\begin{equation} \label{eq:ScalRelRRabGaia}
\begin{split}
\text{For RRab $\star$:}\hspace{0.1cm}\text{Amp}_{\rm los} = 39 \cdot \text{Amp}_{G}^{3} - 133 \cdot \text{Amp}_{G}^{2}  \\
+ 159 \cdot \text{Amp}_{G}.
\end{split}
\end{equation}
\vspace{-0.65cm}
\begin{equation} \label{eq:ScalRelRRcGaia}
\text{For RRc\phantom{b} $\star$:}\hspace{0.1cm}\text{Amp}_{\rm los} = 62(2) \cdot \text{Amp}_{G} \hspace{0.1cm}, \hspace{0.3cm} \varepsilon_{\text{RRc}} = 4.
\end{equation}
\begin{equation}
\begin{split}
\text{Cov}^{\rm RRab} = 
\begin{bmatrix} 
172 & -271 & 97 \\ 
-271 & 438 & -163 \\ 
97 & -163 & 64 \\
\end{bmatrix} \hspace{0.3cm} \varepsilon_{\text{RRab}} = 5.
\end{split} \\
\end{equation}

\begin{figure*}
\includegraphics[width=2\columnwidth]{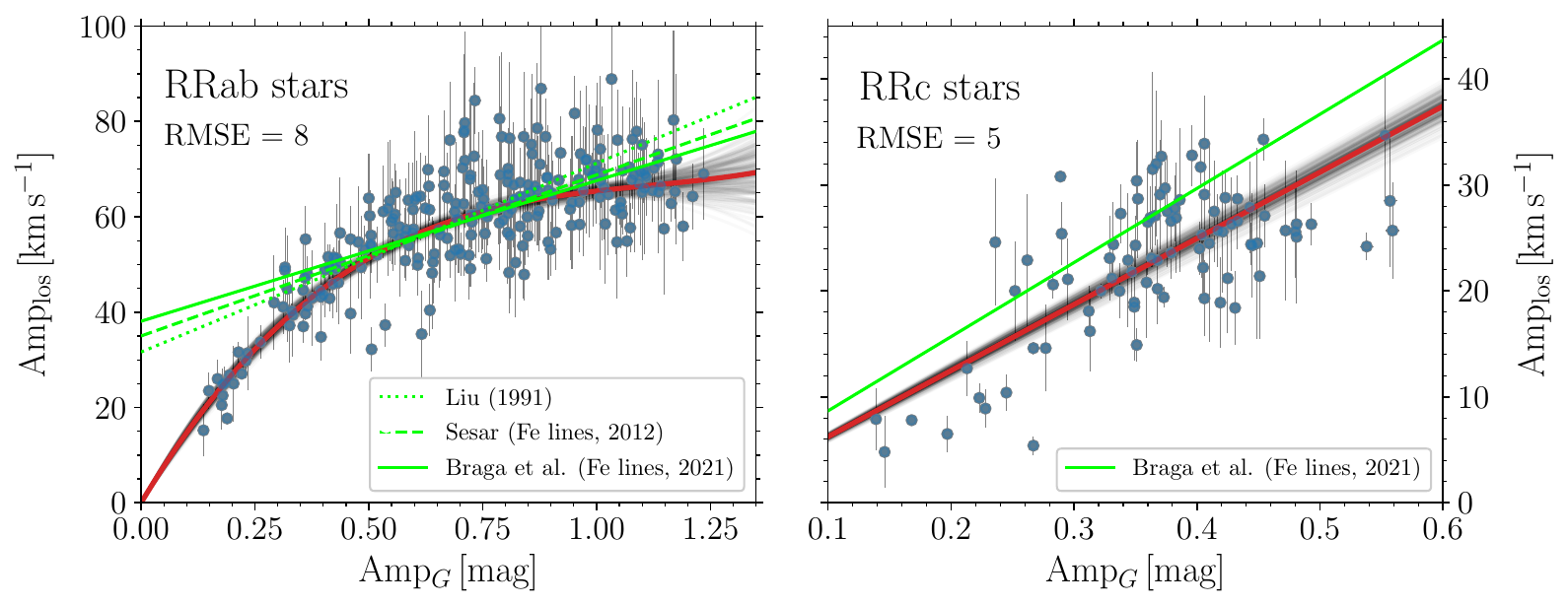}
\caption{Same as Fig.~\ref{fig:ScallingRelaFIG} but for line-of-sight velocity amplitudes determined based on \textit{Gaia} calcium triplet spectra. In addition, the green solid, dashed, and dotted lines represent linear relations from \citet{Braga2021}, \citet{Sesar2012}, and \citet[][using $p$-factor equal to $1.35$]{Liu1991}, respectively. We note that the literature relations were scaled using Eq.~\ref{eq:AmpsConversionVG} and \ref{eq:AmpsConversionVGRRc}.}
\label{fig:ScallingRelGaia}
\end{figure*}

\subsection{Line-of-sight velocity templates for Gaia spectra} \label{subsec:GaiaTemp}

To generate line-of-sight velocity templates from measured $v_{\rm los}$, we proceeded in the same way as for the APOGEE spectra (see Section~\ref{subsec:TempAPOAA}). In addition to the procedure for the APOGEE spectra, thanks to the large number of measurements, we also included a criterion on the maximum acceptable root-mean-square error (RMSE; measured using available data and the GP regressor for each velocity curve) for a given line-of-sight velocity curve in the following form: $\text{RMSE} < 4$. We again normalized the phased line-of-sight velocity curves and decomposed them with Eq.~\ref{eq:FourSerTemp} using the fifth and third degree for the fundamental and first-overtone variables, respectively. Following \citet{Braga2021} and taking advantage of the large dataset, we divided our RRab sample into three groups based on their pulsation periods: RRab-1 (pulsation periods below $0.55$\,day), RRab-2 (pulsation periods above $0.55$\,day and below $0.7$\,day), and RRab-3 (pulsation periods above $0.7$\,day). This division of RRab variables in period bins is based on the arguments presented in \citet[][see their Section~3.1 for details]{Braga2019}. In principle, the photometric amplitudes are not linearly dependent on pulsation periods, which are tightly connected to stellar parameters (e.g., mass, luminosity, metallicity).

In addition to the criteria mentioned in the previous subsection, we selected only the highest quality line-of-sight velocity curves. These include those on the bright end of the mean magnitude distribution ($V < 12.0$\,mag) and those with at least $25$ measurements of $v_{\rm los}$. We made an exception in the case of the RRab-3 category, relaxing the criterion for the number of measurements to at least $20$ measurements of $v_{\rm los}$. Final templates for all four categories can be found in Figure~\ref{fig:tempGaia}, and their Fourier coefficients are listed in Table~\ref{tab:FourTEMPGaia}. As in the case of APOGEE templates, we estimated the RMSE along the templates and list it in Table~\ref{tab:TEMPerrGaia}. The RR~Lyrae variables used for template creation are listed in Table~\ref{tab:FourTEMPGaiaUsedStars}.

\begin{table*}
\caption{Same as Table~\ref{tab:FourTEMPApo} but for Fourier coefficients of the line-of-sight velocity templates for \textit{Gaia} spectra.}
\label{tab:FourTEMPGaia}
\begin{tabular}{lccccccccccc}
\hline
Template  & $a_0$     & $a_1$      & $a_2$      & $a_3$      & $a_4$      & $a_5$     & $a_6$      & $a_7$     & $a_8$      & $a_9$     & $a_{10}$    \\ \hline
$\texttt{Temp}^{\rm RRab-1}$ & $0.0000$ & $-0.2699$ & $-0.2723$ & $-0.0540$ & $-0.1428$ & $0.0120$ & $-0.0955$ & $0.0402$ & $-0.0493$ & $0.0381$ & $-0.0115$ \\
$\texttt{Temp}^{\rm RRab-2}$ & $0.0000$ & $-0.2983$ & $-0.2924$ & $-0.0538$ & $-0.1330$ & $-0.0101$ & $-0.0931$ & $0.0419$ & $-0.0642$ & $0.0511$ & $-0.0206$ \\
$\texttt{Temp}^{\rm RRab-3}$ & $0.0000$ & $-0.2931$ & $-0.2508$ & $-0.0772$ & $-0.1432$ & $0.0057$ & $-0.1009$ & $0.0577$ & $-0.0403$ & $0.0386$ & $-0.0041$ \\
$\texttt{Temp}^{\rm RRc}$  & $0.0000$ & $-0.1997$ & $-0.3880$ & $0.0328$ & $-0.1382$ & $0.0455$ & $-0.0297$ & --      & --       & --      & -- \\ \hline
\end{tabular}
\end{table*}

\begin{figure*}
\includegraphics[width=2\columnwidth]{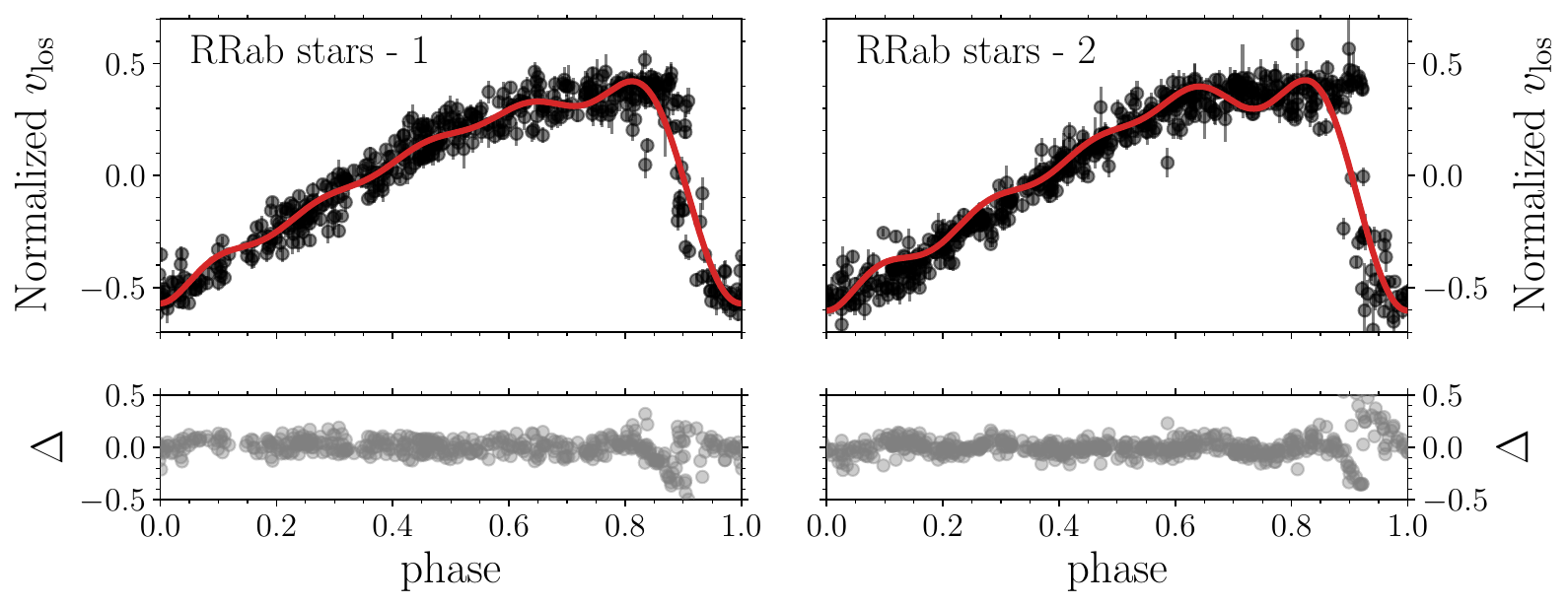}
\includegraphics[width=2\columnwidth]{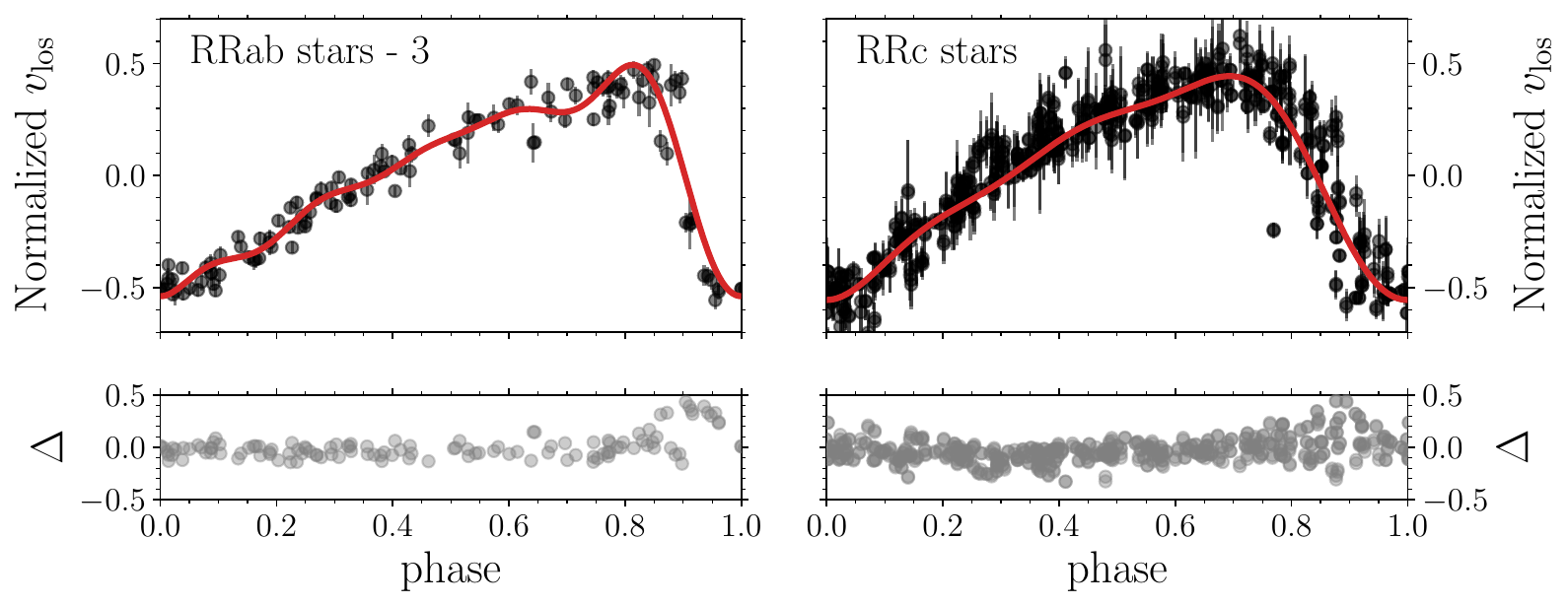}
\caption{Same as Fig.~\ref{fig:tempAPO} but for the line-of-sight velocity curves determined based on \textit{Gaia} calcium triplet spectra.}
\label{fig:tempGaia}
\end{figure*}

\section{Nonlinearity of the scaling relations} \label{sec:NonLinearity}

One unanticipated finding illustrated in Figure~\ref{fig:ScallingRelGaia} is that the scaling between photometric and line-of-sight velocity amplitudes for fundamental mode pulsators is not linear. Instead, a break at $\approx 0.4$\,mag ($\approx 45$\,km\,s$^{-1}$) is apparent, which will primarily affect the shorter period RRab stars. In this section, we attempt to understand the underlying processes governing this behavior through the use of nonlinear pulsation models. 

To compare the nonlinear trend in the RRab scaling relation, we used photometric data from the All-Sky Automated Survey for Supernovae \citep[ASAS-SN;][]{Shappee2014,Jayasinghe2018}. The ASAS-SN survey uses $24$ small telescopes ($14$\,cm in diameter) located all around the globe, continuously scanning the night sky. Photometry in the $V$-band and $g$-band are provided for many RR~Lyrae stars, with typical uncertainties of $0.02$\,mag for stars brighter than $12.0$\,mag in $V$-mag. We used the ASAS-SN $V$-band photometry to establish the basic pulsation properties (pulsation periods, times of brightness maxima, and amplitudes of the light changes) for our RR~Lyrae template dataset. We proceeded in a manner similar to our analysis in Section~\ref{sec:GaiaCaT} by optimizing the Fourier light curve decomposition through Eq.~\ref{eq:FourSer}. For this analysis, we used only RRab stars that were utilized in Section~\ref{subsec:ScalRelGaia}. In addition to the criteria for scaling relations, we used only stars above $b > 10$\,deg. This criterion was selected to reduce the possible blending of stars, which would decrease pulsation amplitudes, particularly since we were using the ASAS-SN photometry, which has large pixel sizes (8 arcsecs).

The pulsation models were computed with the Radial Stellar Pulsation (RSP) tool implemented as part of the Modules for Experiments in Stellar Astrophysics \citep[MESA;][]{Paxton2011,Paxton2013,Paxton2015,Paxton2018,Paxton2019,Jermyn2023}. We used the MESA r21.12.1 public release. The RSP tool computes Lagrangian nonlinear convective pulsation models with time-dependent convection-pulsation coupling given by the \cite{Kuhfuss1986} model. In general, numerical implementation follows the work of \cite{Smolec2008}. First, RSP builds a static equilibrium model. Then, linear stability analysis is conducted, which yields linear pulsation periods and growth rates of the radial modes. Finally, after applying initial perturbation following the line-of-sight velocity eigenvector of a given mode, the model is integrated in time until it approaches a finite amplitude, single-periodic limit cycle pulsations.

We built a chemically homogeneous envelope with $180$ cells for all model computations. Sixty surface cells have an equal mass down to the anchor zone where the temperature is fixed to $11\cdot 10^3$\,K. Then, the mass of the cells increases geometrically inward down to a temperature of  $2\cdot 10^6$\,K. The envelope is nonrotating, with constant luminosity at the radiative, rigid bottom boundary. The OPAL opacities were used \citep{Iglesias1996}, and at lower temperatures, they were supplemented with the \cite{Ferguson2005} opacity data. The distribution of metals followed the solar abundance distribution as given by \cite{Asplund2009}. Our basic model sequences adopted convective parameters of set A, as given in Table~4 of \cite{Paxton2019}.

We were interested in computing fundamental-mode models covering a full spectrum of pulsation amplitudes, starting with those exhibiting the smallest amplitudes, which are expected to be close to the edges of the fundamental-mode instability strip (IS). Therefore, we considered model sequences with a constant mass and luminosity and an increasing effective temperature that start close to the red edge of the fundamental-mode IS. The red edge for the fundamental mode is well beyond the IS for the first overtone, which is located at hotter temperatures. Consequently, the models developed full-amplitude fundamental-mode pulsation, which is the regime of interest for this case. If we constructed our sequence of models near the blue edge of the fundamental-mode IS, the models would also be inside the first-overtone IS. Consequently, near the blue boundary for the fundamental mode IS, the model would switch to pulsations in the first overtone instead of developing pulsations in the fundamental mode. Pulsations in the fundamental mode would be possible for lower temperatures, further from the blue edge of the fundamental mode IS, but then the amplitudes would already be significant. (For more information on the mode selection, see the works of, e.g., \cite{Szabo2004,Smolec2008b, Paxton2019}.)

Altogether, we considered eight model sequences: four with $\mathrm{[Fe/H]}=-1.0$\,dex ($Z=0.0014$, $X=0.75$) and four with $\mathrm{[Fe/H]}=-2.0$\,dex ($Z=0.00014$, $X=0.75$). For each metallicity, the adopted mass is either $0.65\MS$, or $0.75\MS$, and the adopted luminosity is either $45\LS$, or $50\LS$. We used Fourier amplitudes ($A_{1}$) instead of peak-to-peak amplitudes to compare with observational data. While pulsation models reproduce the low-order Fourier parameters of the light curves well, they fail when it comes to sharp features in short wavelengths \citep[see, e.g.,][]{Marconi2015,Das2018,Paxton2019}. In fundamental mode, RR~Lyrae models quite often have sharp spikes that develop close to the maximum light and affect the peak-to-peak amplitude, but these sharp spikes are not observed in the light curves. In Fig.~\ref{fig:ModelsDiffPhys}, we compare the model amplitudes with observations. For the line-of-sight velocity amplitude, we considered photosphere pulsation velocities converted to observed line-of-sight velocities using a fixed projector factor of $1.35$ \citep{Kovacs2003}. The overall agreement between models and observations is very good; in particular, the correlation between the velocity and light curve amplitudes is well reproduced. Theoretically, at low amplitudes, the relation between the photometric and line-of-sight velocity is linear. It then breaks at $\approx 0.2$\,mag but soon becomes linear again, with a shallower slope. The spread of the line-of-sight velocity amplitudes at a given light amplitude may be explained as being due to different physical parameters of the stars (e.g., different masses, luminosities, metallicities). While the overall agreement is satisfactory, we note that the model relation seems to be shifted toward higher photometric amplitudes in a low-amplitude regime; conversely, the predicted line-of-sight velocity amplitude is too small at a given light amplitude. Different projection factors cannot rectify this shift, especially at the low-amplitude regime.

Physical parameters (mass, luminosity, metallicity) are not the only ones to affect light and line-of-sight velocity amplitudes, or more broadly their shape. Parameters of the convection model are also well known to affect the light and line-of-sight velocity curves. To check their effect, for one set of physical parameters ($0.65\MS$, $45\LS$, $\mathrm{[Fe/H]}=-2.0$\,dex), we considered three more model sequences with convective parameters corresponding to those of sets B, C, and D of \citet[][their Table~4]{Paxton2019} in addition to set A, for which previously described model sequences were computed. Sets B and D include the effects of radiative losses, while in sets C and D, the effects of turbulent pressure and kinetic turbulent energy flux are included. The results are illustrated in Fig.~\ref{fig:ModelsDiffConvParam}. We observed that in the low-amplitude regime, the relation between amplitudes is similar for sets A and B and qualitatively different than for sets C and D. For the latter two, the agreement with observations is significantly better, that is, for a given light amplitude, we observed higher line-of-sight velocity amplitudes. The observations in the low-amplitude range are well matched with sets C and D. These two sets include the effects of turbulent pressure and turbulent kinetic energy flux. It seems crucial, however, that the models of low amplitude in sets C and D are significantly cooler than the models of sets A and B. This is because the red edge of the IS, which is where we started the model sequences, shifts toward lower effective temperatures by about $150$\,K for sets C and D. For a higher amplitude range, we observed that the models including radiative losses (sets B and D) lead to lower line-of-sight velocity amplitudes at a given light amplitude than sets that neglect radiative losses. We note that there are no strong theoretical constraints for the parameters of the convective model. While some sets are commonly used in the literature, a comprehensive calibration employing numerous observational constraints (e.g., full shapes of the line-of-sight velocity and multi-band light curves) has been performed only sporadically \citep[e.g.,][]{DiFabrizio2002}. Our results illustrate that simultaneous matching of the light and line-of-sight velocity curves is a possible avenue for calibrating convection models.

\begin{figure}
\includegraphics[width=\columnwidth]{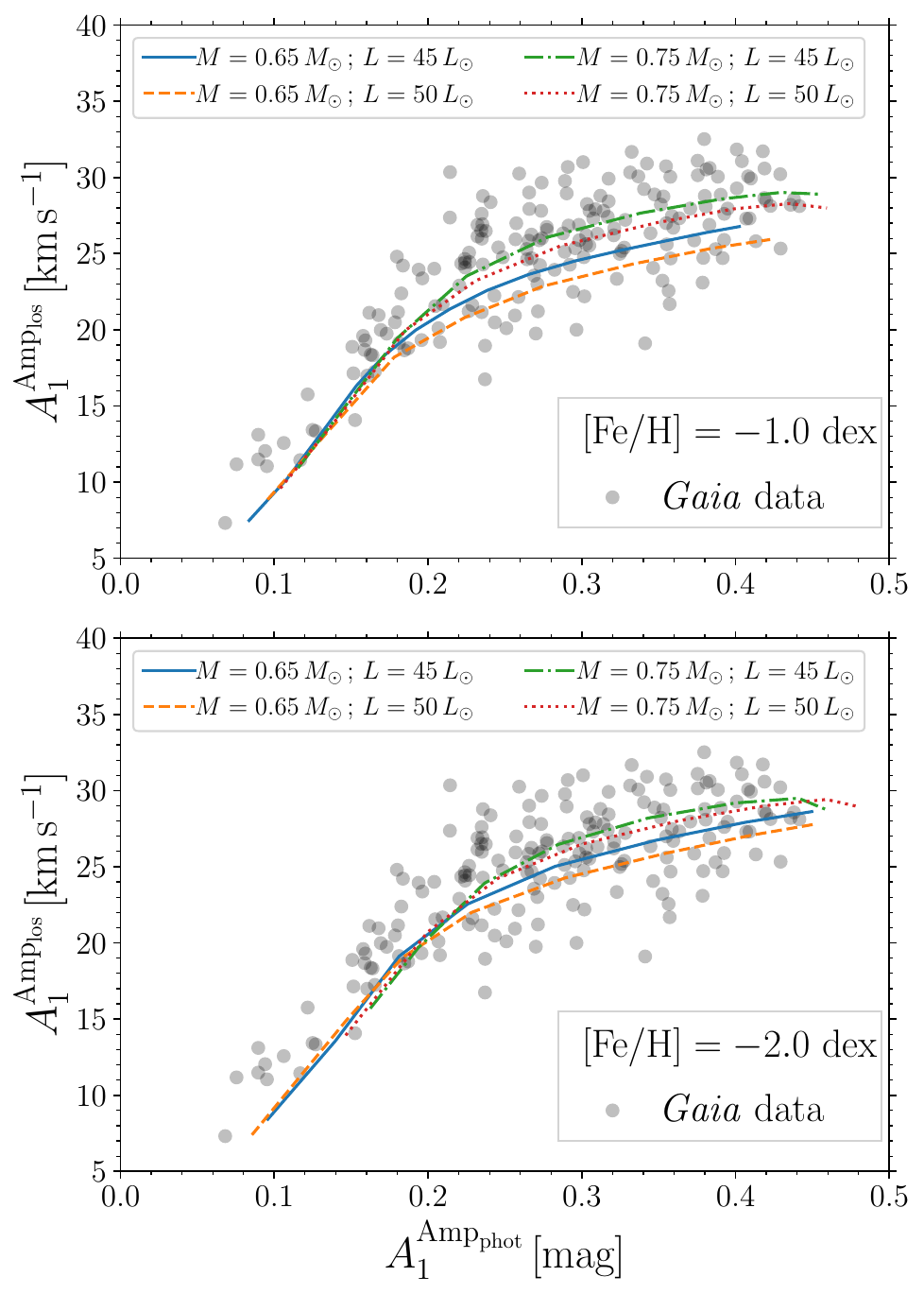}
\caption{Relation between the photometric and line-of-sight velocity Fourier amplitudes for RRab stars (gray dots) along with nonlinear pulsation models of $\mathrm{[Fe/H]}=-1.0$\,dex (top panel) and $\mathrm{[Fe/H]}=-2.0$\,dex (bottom panel) and different masses and luminosities, as indicated in the legend.}
\label{fig:ModelsDiffPhys}
\end{figure}

\begin{figure}
\includegraphics[width=\columnwidth]{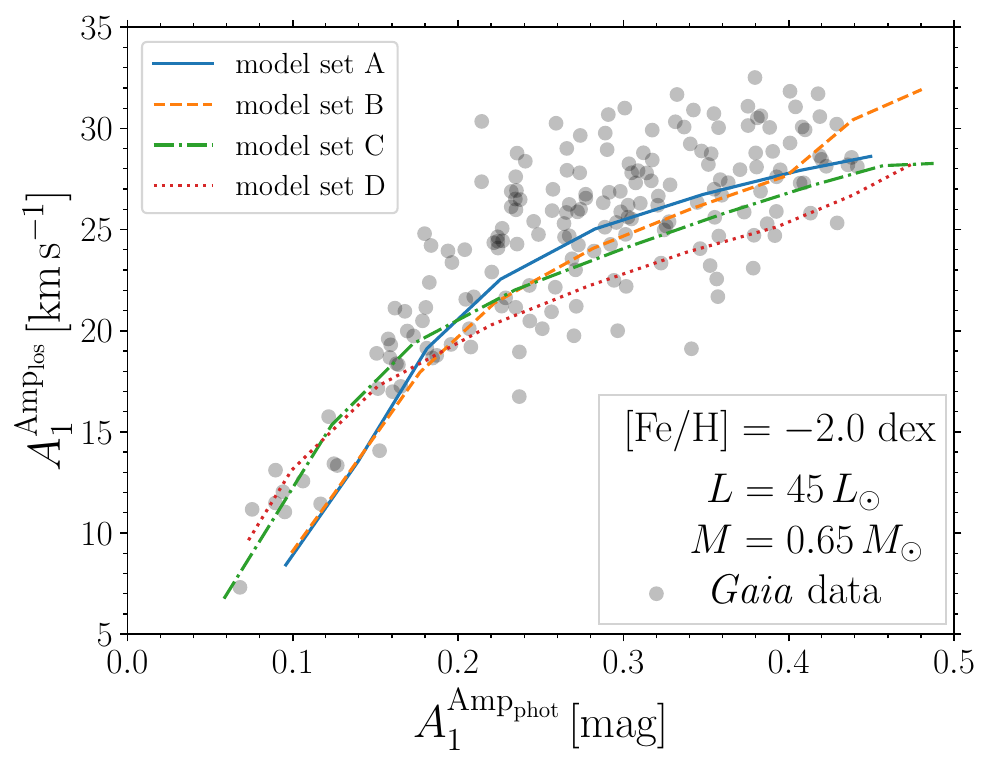}
\caption{Relation between the photometric and line-of-sight velocity Fourier amplitudes for RRab stars (gray dots) along with four sequences of nonlinear pulsation models. We used the following fixed physical parameters ($M=0.65\MS$, $L=45\LS$, $\mathrm{[Fe/H]}=-2.0$\,dex, with varying effective temperature) and assumed different sets of convective parameters.}
\label{fig:ModelsDiffConvParam}
\end{figure}

\section{Cross-survey analysis: APOGEE-\textit{Gaia} velocity templates} \label{sec:bothTemp}

In this section, we explore the feasibility of exchanging derived line-of-sight velocity templates and scaling relations between the APOGEE  ($15000 - 17000$\,\AA) and \textit{Gaia} ($8450 - 8720$\,\AA) surveys. This approach is motivated by the limited number of calibration RR Lyrae stars in the APOGEE survey, specifically both RRab and RRc pulsators. This limitation results in increased scatter across the template. We also evaluate how previously derived templates, particularly from \citet{Braga2021} and \citet{Liu1991}, align with our line-of-sight velocity templates, with a focus on the \citet{Braga2021} template for RRc pulsators. The interchangeability of the line-of-sight velocity curves has been explored in \citet{Clementini2023}, and the authors found that the line-of-sight velocity measurements from the \textit{Gaia} RVS spectra are in good agreement with the line-of-sight velocities in the literature despite the different spectral ranges.}

\begin{figure*}
\includegraphics[width=2\columnwidth]{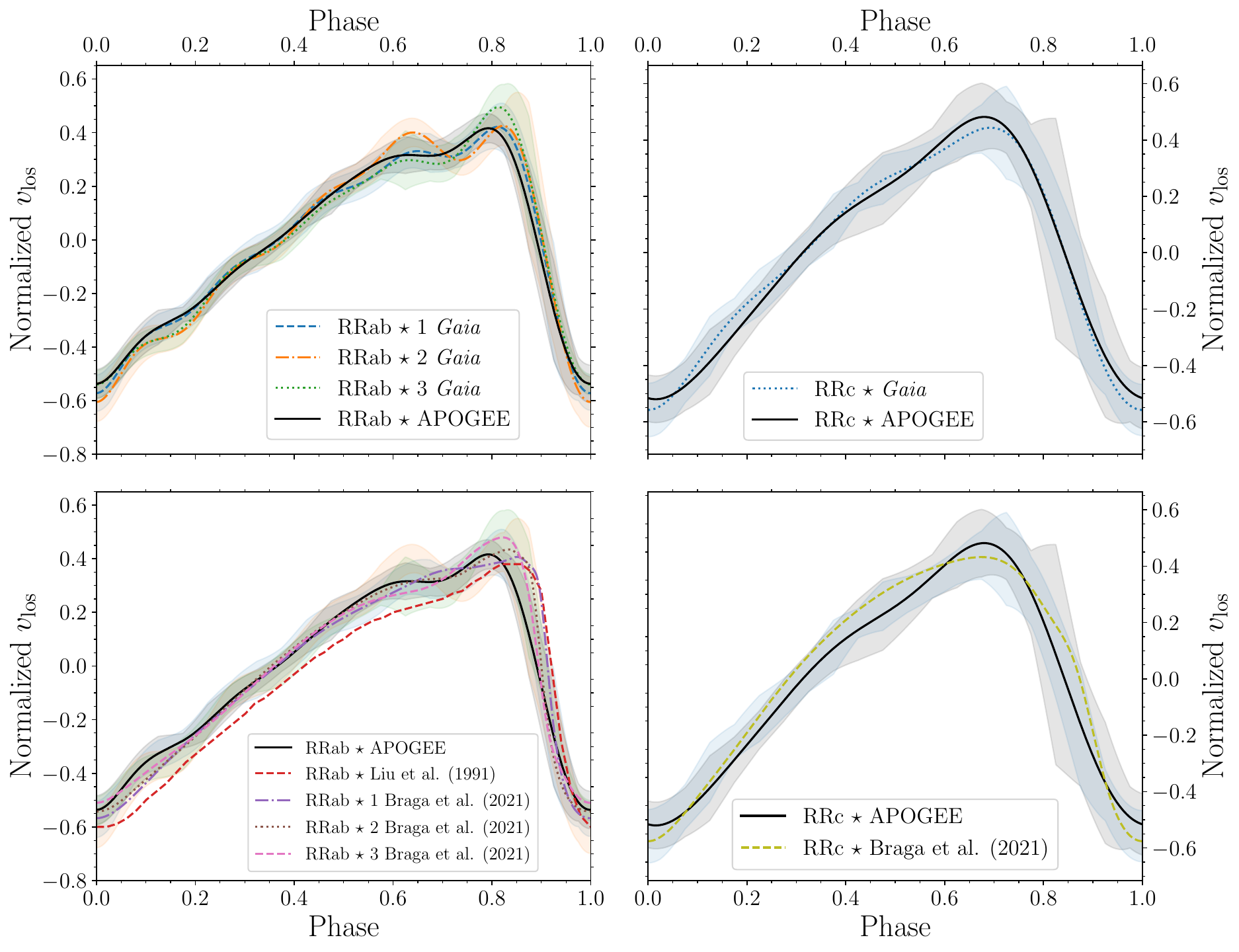}
\caption{Comparison between line-of-sight velocity templates derived in this work (top panels, based on APOGEE and \textit{Gaia} data) and those derived in previous studies (bottom panels, together with APOGEE line-of-sight velocity template). For each line-of-sight velocity template derived in this work, we also included a shaded region (in top and bottom panels) depicting the measured scatter along the template (see Tables~\ref{tab:TEMPerr} and \ref{tab:TEMPerrGaia}).}
\label{fig:tempComp}
\end{figure*}

In Figure~\ref{fig:tempComp}, we present a comparison between the newly derived line-of-sight velocity templates based on APOGEE and Gaia spectra. In addition, we show a comparison with the previously derived templates in \citet{Liu1991} and \citet{Braga2021} for Fe lines. The top panel shows that the line-of-sight velocity templates from APOGEE and Gaia are largely interchangeable, particularly when considering the scatter associated with each template. This indicates that templates developed using Gaia spectra can effectively be employed for determining systemic velocities from APOGEE data. Furthermore, the bottom panel of Figure~\ref{fig:tempComp} illustrates that the line-of-sight velocity templates derived from iron lines in \citet{Braga2021} are also applicable to APOGEE and Gaia data. This compatibility is particularly notable for the RRab templates, where all three template sources (APOGEE, Gaia, and \citet{Braga2021}) exhibit remarkably similar patterns. The only notable deviation occurs with the RRab template from \citet{Liu1991}, which shows an offset that is also in phase; adjusting it by $-0.05$ would align it closely with the APOGEE template. This comparative analysis of the templates implies that various sources for the line-of-sight velocity templates are effectively interchangeable within the errors.

We focus next on the substitutability of the scaling relations between surveys and the spectral lines used in the determination of line-of-sight velocities. Previously, in Fig.~\ref{fig:ScallingRelGaia}, we demonstrated that the scaling relations for the RRab variables from \citet{Liu1991} and \citet{Braga2021} accurately represent the higher amplitude regime. For a specific RRab star with a given photometric amplitude of Amp$_{G} = 0.8$\,mag (Amp$_{V} = 1.06$\,mag, using Eq.~\ref{eq:AmpsConversionVG}), the line-of-sight velocity amplitudes calculated using the relations from \citet{Liu1991}, \citet{Sesar2012}, \citet{Braga2021}, and Eq.~\ref{eq:ScalRelRRabGaia} fall within the range of $61$ to $62$ km/s. Using the scaling relation for RRab stars with APOGEE data (see Eq.~\ref{eq:ScalRel}) yields a higher Amp$_{\rm los} = 61$\,km\,s$^{-1}$.

A similar conclusion can be drawn for RRc stars. We compared the amplitude scaling relations derived for the RRc stars based on \textit{Gaia}, APOGEE, and the \citet{Braga2021} relation (based on iron lines of first-overtone pulsators). This comparison revealed that a typical RRc with an Amp$_{G} = 0.35$\,mag (Amp$_{V} = 0.47$\,mag, using Eq.~\ref{eq:AmpsConversionVGRRc}) has a predicted Amp$_{\rm los}$ equal to $22$, $22$, and $28$\,km\,s$^{-1}$ for relations from \citet{Braga2021}, Eq.~\ref{eq:ScalRel}, and Eq.~\ref{eq:ScalRelRRcGaia}, respectively. Our predictions suggest a slightly lower Amp$_{\rm los}$ in comparison with the prediction by \citet{Braga2021}. There could be several reasons for this, especially the differences in covered amplitude space. Our \textit{Gaia} dataset covers both a higher and lower Amp$_{G}$ range than \citet{Braga2021}. In the APOGEE dataset, we do not have as many data points as were used in \citet{Braga2021} and in our \textit{Gaia} dataset.

The comparisons outlined above demonstrate the possibility of substituting templates and scaling relations across surveys. For example, one can use line-of-sight velocity templates and scaling relations based on \textit{Gaia} spectra to derive systemic velocities for line-of-sight velocities estimated from APOGEE spectra. Despite differences in wavelength coverage and spectral lines between the surveys, the estimated systemic velocities using either method are comparable, as long as we are not in a low-amplitude regime for RRab stars. This yields an advantage, especially for RRab stars that fall into the low-amplitude space, where we can use the well-described scaling relation based on the \textit{Gaia} dataset across different line-of-sight velocity sources. One can even use the \textit{Gaia} line-of-sight velocity templates for velocities calculated based on iron lines. In principle, all the variations should lead to equivalent predicted Amp$_{\rm los}$ and systemic velocities within the uncertainties.

\section{An example of determining $v_{\rm sys}$ and testing the newly derived templates} \label{sec:ExampleOfAnalysisAndTesting}

In what follows, we describe a method for estimating systemic velocities for a selected RR~Lyrae variable using the derived templates and scaling relations presented above. We assess the resulting systemic velocity and apply our method to the large portion of the \textit{Gaia} RR~Lyrae catalog with $v_{\rm los}$.\footnote{We estimated $v_{\rm sys}$ using our templates and scaling relations only for RR~Lyrae stars in the \textit{Gaia} catalog that passed the criteria outlined in Subsection~\ref{subsec:ScalRelGaia}.} 

\subsection{An example of determining $v_{\rm sys}$ using APOGEE data} \label{subsec:ExampleOfAnalysis}

Most of the RR~Lyrae stars observed by the APOGEE survey have well-known pulsation properties (pulsation period, amplitude, and time of brightness maxima) but only a few spectroscopic observations ($1-3$), so the coverage of the pulsation cycle is incomplete. In cases similar to this, the line-of-sight velocity template and scaling relation need to be used to derive the $v_{\rm sys}$ and $\sigma_{v_{\rm sys}}$. For this exercise, we chose BH~Aur, which belongs to the RRab subclass, with $P = 0.4560898$\,day, Amp$_{G} = 0.81 \pm 0.03$\,mag, and $E = 2456907.7783865286$\,day (based on \textit{Gaia} photometry). It has only four APOGEE subvisit observations and a well-determined $v_{\rm sys} = 51.4 \pm 1.5$\,km\,s$^{-1}$ in the \textit{Gaia} survey\footnote{Determined by the \textit{Gaia} survey using the Fourier decomposition of the line-of-sight velocity curve for BH~Aur.} due to a large number of spectroscopic observations \citep[$28$ in total, see \textit{Gaia} RR~Lyrae catalog][]{Clementini2023}.

In the determination of $v_{\rm sys}$ and $\sigma_{v_{\rm sys}}$, we first converted the photometric amplitude in the $G$-band to the $\text{Amp}_{\rm los}$ using the relation in Eq.~\ref{eq:ScalRel} and estimated the uncertainty on the $\text{Amp}_{\rm los}$, $\sigma_{\text{Amp}_{\rm los}}$. This allowed us to assemble a dataset $\mathbf{D}$ for BH~Aur that consisted of its pulsation properties, measured $v_{\rm los}$ (including their associated uncertainties), and the time of observation represented by the Barycentric Julian Date (based on APOGEE subvisit spectra):
\begin{equation}
\mathbf{D} =\left\{ P, E, \text{Amp}_{\rm los}, \sigma_{\text{Amp}_{\rm los}}, v_{\rm los}, \sigma_{v_{\rm los}}, T_{\rm BJD}^{\rm APO} \right \} \\.
\end{equation}
To find $v_{\rm sys}$ and its uncertainty given the dataset $\mathbf{D}$, we optimized a subsequent relation for likelihood $p(\mathbf{D}\,|\,v_{\rm sys})$ defined as
\begin{equation} \label{eq:vsysAPOlike}
p(\mathbf{D}\,|\,v_{\rm sys}) = \mathcal{N}\left( v^{\rm model}_{\rm los} \,|\,v_{\rm los}, \sigma^{\rm model}_{\rm data} \right) \\,
\end{equation}
where $\mathcal{N}$ represents the normal distribution. The $v^{\rm model}_{\rm los}$ and $\sigma^{\rm model}_{\rm data}$ are described by two following equations:
{\footnotesize
\begin{gather} \label{eq:vlos_model}
v^{\rm model}_{\rm los} = \text{Amp}_{\rm los}\cdot\texttt{Temp}^{\rm RR} \left( \Phi \right) + v_{\rm sys} \\
\sigma^{\rm model}_{\rm data} = \sqrt{\sigma_{v_{\rm los}}^{2} + \left (\texttt{Temp}_{\text{err}}^{\rm RR} \left ( \Phi \right )\cdot \text{Amp}_{\rm los} \right )^{2} + \left( \texttt{Temp}^{\rm RR} \left( \Phi \right) \cdot \sigma^{\rm model}_{\rm data} \right )^{2} } .
\end{gather}}
In these equations, $\Phi^{k}$ stands for a pulsation phase of the observation calculated using the BH~Aur ephemerides and $T_{\rm HJD}^{\rm APO}$ in the following relation:
\begin{equation}
\Phi = \texttt{mod} \left( T_{\rm BJD}^{\rm APO} - M_{\rm 0}, P\right) / P \\.
\end{equation}
The terms $\texttt{Temp}^{\rm RR}$ and $\texttt{Temp}_{\text{err}}^{\rm RR}$ represent the line-of-sight-velocity template (described by the Fourier series; see first row in Table~\ref{tab:FourTEMPApo}) and a spline function that describes the scatter in the template\footnote{The uncertainty in the template ($\texttt{Temp}_{\rm err}^{\rm RR}$) was estimated as the scatter between the Fourier model and the phased line-of-sight velocity measurements of the template stars along the template (see Sec.~\ref{subsec:TempAPOAA}). (see Table~\ref{tab:TEMPerr}).}

Following the steps above and applying one of the likelihood minimalization (or log-likelihood maximization) routines \citep[e.g., \texttt{SciPy} and \texttt{emcee} ][]{scipy,Foreman-Mackey2013}, we estimated $v_{\rm sys} = 49.7 \pm 1.6$\,km\,s$^{-1}$ for BH~Aur.\footnote{An example of the described analysis can be found in the GitHub repository \url{https://github.com/ZdenekPrudil/Vlos-templates}} Our derived value based on APOGEE spectra is in agreement with the systemic velocity estimated by \textit{Gaia}, $v_{\rm sys} = 51.4 \pm 1.5$\,km\,s$^{-1}$. 

\subsection{Testing the newly derived scaling relations and templates} \label{subsec:TestTempGaia}

In this subsection, we assess the accuracy of the derived line-of-sight velocity templates by estimating the systemic velocities for the large fraction of the \textit{Gaia} RR~Lyrae RVS dataset. In order to compile a sample of RR~Lyrae stars with $v_{\rm sys}$ to use as labels (i.e., the "ground truth"), we used only single-mode RR~Lyrae stars and criterion on \texttt{ipd\_frac\_multi\_peak}~$< 5$ to avoid any blended stars. This resulted in a sample of $1083$ RR~Lyrae stars, and these stars were used to examine the precision of the new scaling relations and the derived templates and their impact on the systemic velocities for RR~Lyrae stars with $v_{\rm los}$ measurements in the \textit{Gaia} catalog. For the estimation of $v_{\rm sys}$, we proceeded in the same fashion as in the previous subsection. We used the determined pulsation properties from the \textit{Gaia} catalog and obtained the systemic velocities for the tested stars \citep{Clementini2023}. 

Since some of the RR~Lyrae variables in the \textit{Gaia} catalog have spurious values for $v_{\rm los}$ that would negatively affect the $v_{\rm sys}$ determination, we imposed a criterion on the $v_{\rm los}$ for a given star. First, for a given RR~Lyrae star, we estimated the median $v_{\rm los}$ value $\left| v_{\rm los} \right| $. The value of $\left| v_{\rm los} \right| $ falls close to the actual $v_{\rm sys}$; thus, in the $v_{\rm sys}$ determination, we only considered $v_{\rm los}$ values falling within $\pm 50$\,km\,s$^{-1}$ of the $\left| v_{\rm los} \right| $. This cut removed nearly all spurious measurements, and the interval of $\pm 50$\,km\,s$^{-1}$ was motivated by the maximum line-of-sight velocity amplitudes (see Fig.~\ref{fig:ScallingRelGaia}), where all $\text{Amp}_{\rm los}$ are smaller than $90$\,km\,s$^{-1}$.

\begin{figure}
\includegraphics[width=\columnwidth]{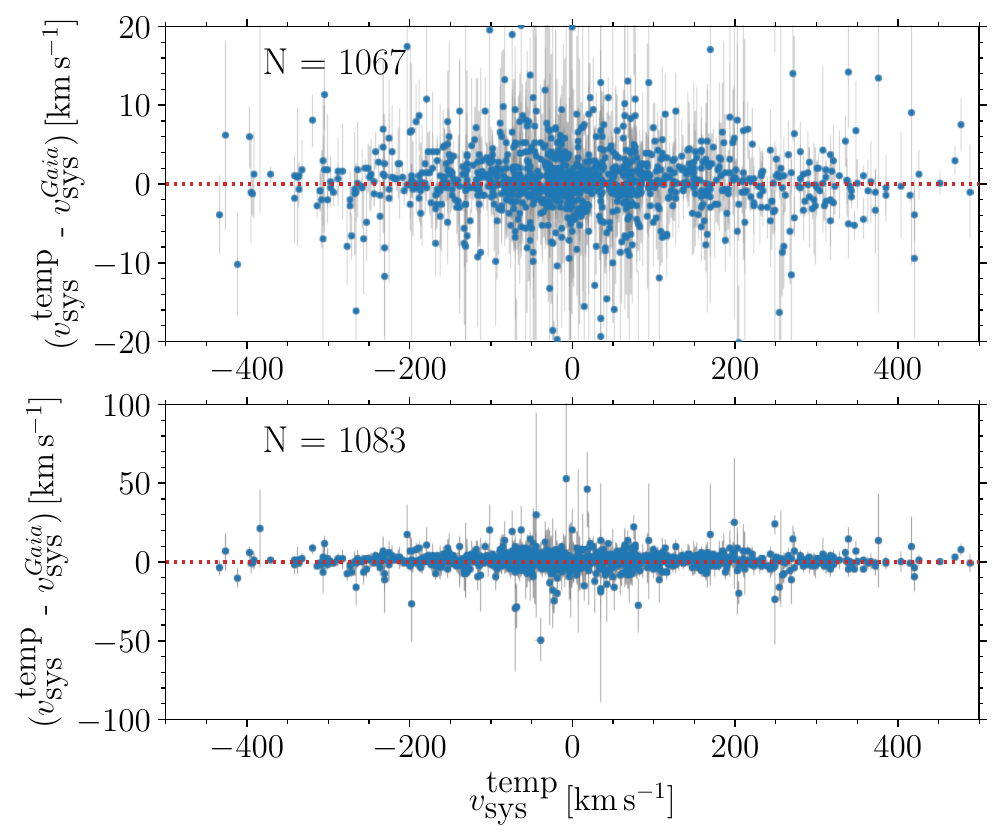}
\caption{Comparison between \textit{Gaia}-determined systemic velocities \citep{Clementini2023} and our determined $v_{\rm sys}$ using newly derived line-of-sight velocity templates and scaling relations. The top panel shows a zoom-in of the stars with the difference between \textit{Gaia} and our values below $|20|$\,km\,s$^{-1}$, while the bottom panel shows the entire scale. The numbers in each panel denote the total number of stars displayed in a given plot.}
\label{fig:ComparisonGaiaAndTemp}
\end{figure}

In Fig.~\ref{fig:ComparisonGaiaAndTemp}, we depict a comparison between our derived values and the $v_{\rm sys}$ measurements in the \textit{Gaia} RR~Lyrae catalog. We found an excellent match for our estimated values and a very low scatter, demonstrating a low RMSE equal to $5.7$\,km\,s$^{-1}$. Several of the outliers in the bottom panel of Fig.~\ref{fig:ComparisonGaiaAndTemp} ($16$ variables with differences above $|20|$\,km\,s$^{-1}$) can be explained by spurious \textit{Gaia} $v_{\rm los}$ measurements that although accompanied by a large $\sigma_{v_{\rm sys}}$ in the \textit{Gaia} RR~Lyrae catalog, still contribute to and influence the \textit{Gaia} $v_{\rm sys}$. Since we tried removing such measurements, our derived values differ from those published in the \textit{Gaia} RR~Lyrae catalog. An example of an RR~Lyrae with spurious measurements is the \texttt{source\_id}~$=1826975050760883072$ (HR~Vul). This star has $v_{\rm los}$ measurements ranging from $823$\,km\,s$^{-1}$ to $-480$\,km\,s$^{-1}$, thus far exceeding the amplitude variation for a classical RR~Lyrae star. For HR~Vul, we found $v_{\rm sys} = -8 \pm 2$\,km\,s$^{-1}$, while the \textit{Gaia} RR~Lyrae catalog lists $v_{\rm sys} = -60 \pm 50$\,km\,s$^{-1}$. The general \textit{Gaia} \texttt{gaia\_source} catalog provides a value for HR~Vul of $v_{\rm sys} = -4.59 \pm 4.35$\,km\,s$^{-1}$.\footnote{This value was determined through a combination of obtained cross-correlation functions for each observed epoch with a single cross-correlation function that was used to estimate $v_{\rm sys}$ and its uncertainty.} This large discrepancy was caused by several spurious measurements of $v_{\rm los}$ that were included in the estimation of $v_{\rm sys}$ for the \textit{Gaia} RR~Lyrae catalog.

The \textit{Gaia} RR~Lyrae catalog provides numerous measurements of $v_{\rm los}$ that, particularly for RR~Lyrae stars in the Galactic bulge, are generally hard to come by. Therefore, we performed a more qualitative test of our method as well as a comparison with $v_{\rm sys}$ values derived by a method used in the past for Galactic bulge RR~Lyrae stars \citep[using line-of-sight velocity template and scaling relation from][]{Liu1991}. We proceeded in the same way as in the case above and only removed first-overtone pulsators since \citet{Liu1991} is calibrated only on fundamental-mode RR~Lyrae stars. To simulate possible observations for RR~Lyrae stars toward the Galactic bulge, we used data published by the BRAVA-RR survey \citep{Kunder2016}, where most variables were observed with four epochs. We selected the same four epochs for individual stars and applied our methods and those of \citet{Liu1991} to compare the two approaches. 

The resulting comparison was made based on the RMSE between \textit{Gaia} $v_{\rm sys}$ values (used as reference values) and values derived using our new method and the procedure described by \citet{Liu1991}. Based on $884$ tested RRab variables, we found a slightly lower RMSE for our newly derived approach, $\text{RMSE}^{\text{\citep{Liu1991}}} = 9.6$\,km\,s$^{-1}$ and $\text{RMSE}^{\text{this work}} = 7.5$\,km\,s$^{-1}$. Furthermore, when we included only bright variables (mean apparent $G$-band magnitudes below $12.0$\,mag) that have very precise $v_{\rm los}$ in the \textit{Gaia} RR~Lyrae catalog in the comparison, we found $\text{RMSE}^{\text{\citep{Liu1991}}} = 5.8$\,km\,s$^{-1}$ and $\text{RMSE}^{\text{this work}} = 4.0$\,km\,s$^{-1}$ for the method by \citet{Liu1991}. Thus, for very precise data, our approach leads to an RMSE smaller by a factor of $1.5$ than the RMSE from the method outlined by \citet{Liu1991}.

Furthermore, we explored how well both methods estimate uncertainties on $v_{\rm sys}$ concerning the selected "ground truth" in \textit{Gaia} $v_{\rm sys}$ values. We first estimated the difference between the $v_{\rm sys}$ values derived through two tested methods and the values from the \textit{Gaia} RR~Lyrae catalog, $\Delta v_{\rm sys}$. Then, for each method, we added uncertainties in quadrature with uncertainties values from \textit{Gaia}, resulting in $\sigma^{\text{tot}}$. Under the assumption that the difference between the selected method and the \textit{Gaia} values follows the normal distribution, we expected to find $\approx68.2$ percent of tested stars within one $\sigma^{\text{tot}}$ of the difference distribution. Thus, from $884$ tested RRab variables, $603$ pulsators should be within one $\sigma^{\text{tot}}$ of the $\Delta v_{\rm sys}$ distribution. When compared with two tested methods, we found that our method and that of \citet{Liu1991} yield $597$ and $466$ pulsators within one $\sigma^{\text{tot}}$, respectively. Therefore, our estimated uncertainties quite well represent the underlying error distribution. Viable room for further improvement in our approach would be to include a term for the uncertainty in the star's ephemerides (e.g., an offset in phase $\Phi$). This term would consider the possible time shifts between photometrically obtained ephemerides and spectral observations. Adding a term for the offset in $E$ could potentially be helpful for RR~Lyrae stars with modulated light curves due to the Blazhko effect, where additional modulation can cause shifts in the phase \citep[e.g., RS~Boo, see Fig.~6 in ][]{Jones1988}. On the other hand, when we compared our results with those from the method of \citet{Liu1991}, we found that it underestimates the uncertainties of $v_{\rm sys}$. One of the conceivable reasons can be due to neglecting the scatter in the template, which is included in our procedure.




\section{Summary} \label{sec:Summary}

In this study, we derived, for the first time, scaling relations between photometric and line-of-sight velocity amplitudes together with line-of-sight velocity templates for visual and near-infrared spectra (covering Ca\,T and $H$ passband). The aforementioned tools can be used for estimating systemic velocities of single-mode RR~Lyrae stars. The main purpose of this study was to derive tools that can be subsequently utilized in studies focused on the structure and kinematics of the Galactic bulge stellar component. In particular, our aim is to provide methods and tools for processing spectroscopic data for RR~Lyrae stars in surveys such as the APOGEE and BRAVA-RR \citep{Zasowski2017,Kunder2016}.

To create scaling relations and line-of-sight velocity templates for the APOGEE survey, we utilized several RR~Lyrae variables in the Solar neighborhood. Due to the subvisit spectra, we were able to cover the entire line-of-sight velocity curve for several RR~Lyrae stars and construct scaling relations and templates for RRab and RRc subclasses. For scaling equations, we chose to use linear relations without the intercept in order to preserve the physical meaning of the relation. We tested and verified our developed approach for APOGEE spectra on one of the nearby RR~Lyrae stars with a precisely determined systemic velocity in the literature.

For the \textit{Gaia} spectra that cover Ca\,T, we proceeded in nearly the same way as in the APOGEE analysis. Thanks to the abundance and precise data provided by the \textit{Gaia} RR~Lyrae catalog, we found that the scaling relation for the fundamental-mode RR~Lyrae stars is nonlinear and requires a polynomial of a higher degree to accurately describe the transformation of the photometric amplitudes into line-of-sight velocity amplitudes. We did not observe such behavior for first-overtone pulsators. The found nonlinearity was further explored using stellar pulsation models implemented in the MESA code. Using Fourier amplitudes, we were able to replicate the trend for spectroscopic and photometric data. It is important to note that we observed a shift toward larger amplitudes in the low-amplitude regime. The offset in low amplitudes cannot be explained by different physical parameters nor by a varying projection factor. To rectify this shift, we tested different sets of convective parameters, and we found that sets C and D that include turbulent pressure and turbulent kinetic energy flux (see Section~\ref{sec:NonLinearity} for details) significantly better describe the low-amplitude space. The effect may be caused by a shift of the IS toward lower effective temperatures. In the low-amplitude regime, the models are significantly cooler than those of sets A and B.

In our analysis, we examined the feasibility of using APOGEE and Gaia survey data interchangeably, focusing on RRab and RRc pulsators. This was driven by the scarcity of calibration RR~Lyrae stars in APOGEE. Our comparisons of new and existing templates found a high degree of compatibility for both single-mode RR~Lyrae types. Furthermore, we evaluated scaling relations for velocity amplitudes derived in our study and for those in the literature. We conclude that despite the differences between the surveys, they provide similar results, thus enhancing the use and interchangeability of \textit{Gaia} and APOGEE to derive templates and scaling relations. 

We have demonstrated that the templates and scaling relations we provided for RR Lyrae stars in our analysis are largely interchangeable with those found in the existing literature \citep[e.g.,][]{Braga2021}, offering a high degree of compatibility. Crucially, our approach distinguishes itself by offering a robust error treatment, notably enhancing the predictability and accuracy of line-of-sight amplitudes for low-amplitude RR Lyrae stars.


We also tested the newly obtained procedure on \textit{Gaia} Ca\,T spectra, and our method accurately estimates systemic velocities for RR~Lyrae stars in the \textit{Gaia} RR~Lyrae catalog \citep{Clementini2023}. Furthermore, we compared our approach for calculating systemic velocities with methods used in the past, and we found that our approach provides an approximately $50$ percent higher precision in the determination of systemic velocities for observations similar to the BRAVA-RR survey strategy. In addition, our approach reliably estimates uncertainties in systemic velocities when compared to the residual distribution between our values and the original \textit{Gaia} dataset.

Obtaining high-quality spectra of RR~Lyrae stars at the distance of the bulge, even when using only moderate resolutions, is an expensive endeavor. Looking toward the future, Rubin-LSST will uncover hundreds of thousands of RR~Lyrae stars out to distances more than $50$ times the distance to the bulge, so an efficient way of obtaining line-of-sight velocity measurements of distant RR~Lyrae stars will also be valuable.  The line-of-sight velocity templates and scaling relations provided here allow the systemic velocity of an RR~Lyrae star to be accurately found from a single line-of-sight measurement at any point during the pulsation cycle, assuming the stars amplitude, period, and time of maximum brightness are known. As such, our templates open up more precise avenues to carry out large kinematic studies of the bulge RR~Lyrae star population as well as beyond. The derived methods for obtaining systemic velocities of RR~Lyrae stars can be utilized by the upcoming large spectroscopic surveys such as 4MOST, WEAVE, and MOONS \citep{4MOST2014,WEAVE2014,MOONS2014}. We will be using our derived scaling relations and templates in forthcoming papers to examine the structure and kinematics of the Galactic bulge in both the northern and southern hemispheres.

\begin{acknowledgements}
RS was supported by the Polish National Science Centre, SONATA BIS grant, 2018/30/E/ST9/00598.
This work has made use of data from the European Space Agency (ESA) mission
{\it Gaia} (\url{https://www.cosmos.esa.int/gaia}), processed by the {\it Gaia}
Data Processing and Analysis Consortium (DPAC,
\url{https://www.cosmos.esa.int/web/gaia/dpac/consortium}). Funding for the DPAC
has been provided by national institutions, in particular the institutions participating in the {\it Gaia} Multilateral Agreement.

Funding for the Sloan Digital Sky Survey IV has been provided by the Alfred P. Sloan Foundation, the U.S. Department of Energy Office of Science, and the Participating Institutions. 

SDSS-IV acknowledges support and resources from the Center for High Performance Computing  at the University of Utah. The SDSS website is www.sdss4.org.

SDSS-IV is managed by the Astrophysical Research Consortium for the Participating Institutions of the SDSS Collaboration including the Brazilian Participation Group, the Carnegie Institution for Science, Carnegie Mellon University, Center for Astrophysics | Harvard \& Smithsonian, the Chilean Participation Group, the French Participation Group, Instituto de Astrof\'isica de Canarias, The Johns Hopkins University, Kavli Institute for the Physics and Mathematics of the Universe (IPMU) / University of Tokyo, the Korean Participation Group, Lawrence Berkeley National Laboratory, Leibniz Institut f\"ur Astrophysik Potsdam (AIP),  Max-Planck-Institut f\"ur Astronomie (MPIA Heidelberg), Max-Planck-Institut f\"ur Astrophysik (MPA Garching), Max-Planck-Institut f\"ur Extraterrestrische Physik (MPE), National Astronomical Observatories of China, New Mexico State University, New York University, University of Notre Dame, Observat\'ario Nacional / MCTI, The Ohio State 
University, Pennsylvania State University, Shanghai Astronomical Observatory, United Kingdom Participation Group, Universidad Nacional Aut\'onoma de M\'exico, University of Arizona, University of Colorado Boulder, University of Oxford, University of Portsmouth, University of Utah, University of Virginia, University of Washington, University of Wisconsin, Vanderbilt University, and Yale University.

This research made use of the following Python packages: \texttt{Astropy} \citep{astropy2013,astropy2018}, \texttt{emcee} \citep{Foreman-Mackey2013}, \texttt{IPython} \citep{ipython}, \texttt{Matplotlib} \citep{matplotlib}, \texttt{NumPy} \citep{numpy}, \texttt{scikit-learn} \citep{Pedregosa2012}, and \texttt{SciPy} \citep{scipy}.
\end{acknowledgements}

\bibliographystyle{aa}
\bibliography{biby} 

\begin{appendix}

\section{Additional tables}

\begin{table*}[]
\caption{List of nearby RR~Lyrae variables observed by APOGEE and marked with the \texttt{rrlyr} label in APOGEE data products. The first two columns denote the APOGEE and alternative identifier, respectively. Columns three and four list the equatorial coordinates, and the last two columns contain $H$-band magnitudes \citep[from \textit{Two-Micron Sky Survey}, 2MASS,][]{Cutri2003,Skrutskie2006} and the number of visits per star. We note that DT-Gem is not included in this table, although it is noted in the APOGEE files; it does not have an APOGEE-ID; and NVISITS is zero for this variable. The asterisk at Alt.~ID indicates a star used to create the line-of-sight velocity templates.}
\label{tab:30RR}
\begin{tabular}{lcrrrrl}
\hline
APOGEE-ID & Alt.~ID & R.A. & DEC & $H$\,[mag] & NVISITS & Type \\ \hline
\href{https://dr17.sdss.org/sas/dr17/apogee/spectro/redux/dr17/stars/apo1m/rrlyr/apStar-dr17-2M05120427+3357469.fits}{2M05120427+3357469} & BH-Aur & 78.017784 & 33.963054   & 10.008 & 1 & RRab      \\
\href{https://dr17.sdss.org/sas/dr17/apogee/spectro/redux/dr17/stars/apo1m/rrlyr/apStar-dr17-2M03115210-1121140.fits}{2M03115210-1121140} & SV-Eri & 47.967082 & -11.353908 & 8.645  & 1 & RRab        \\
\href{https://dr17.sdss.org/sas/dr17/apogee/spectro/redux/dr17/stars/apo1m/rrlyr/apStar-dr17-2M06224432+1831533.fits}{2M06224432+1831533} & V0397-Gem            & 95.684707 & 18.531471  & 10.779 & 1 & RRc        \\
\href{https://dr17.sdss.org/sas/dr17/apogee/spectro/redux/dr17/stars/apo1m/rrlyr/apStar-dr17-2M06305818+3831220.fits}{2M06305818+3831220} & NSVS-4568675         & 97.741958 & 38.522778  & 10.283 & 1 & RRab        \\
\href{https://dr17.sdss.org/sas/dr17/apogee/spectro/redux/dr17/stars/apo1m/rrlyr/apStar-dr17-2M07113502+4046370.fits}{2M07113502+4046370} & TZ-Aur               & 107.895919   & 40.776974  & 10.761 & 1 & RRab        \\
\href{https://dr17.sdss.org/sas/dr17/apogee/spectro/redux/dr17/stars/apo1m/rrlyr/apStar-dr17-2M07124566+4025223.fits}{2M07124566+4025223} & CSS-J071245.6+402522 & 108.190246 & 40.422916  & 11.052 & 1 & CST$\dagger$      \\
\href{https://dr17.sdss.org/sas/dr17/apogee/spectro/redux/dr17/stars/apo1m/rrlyr/apStar-dr17-2M07333173+4748098.fits}{2M07333173+4748098} & TV-Lyn               & 113.382209 & 47.802723  & 10.414 & 1 & RRc        \\
\href{https://dr17.sdss.org/sas/dr17/apogee/spectro/redux/dr17/stars/apo1m/rrlyr/apStar-dr17-2M07450630+4306415.fits}{2M07450630+4306415} & TW-Lyn               & 116.276161 & 43.111584  & 10.797 & 1 & RRab        \\
\href{https://dr17.sdss.org/sas/dr17/apogee/spectro/redux/dr17/stars/apo1m/rrlyr/apStar-dr17-2M07461785+4424185.fits}{2M07461785+4424185} & GSC-02971-01335      & 116.574409 & 44.405140  & 10.447 & 1 & RRc       \\
\href{https://dr17.sdss.org/sas/dr17/apogee/spectro/redux/dr17/stars/apo1m/rrlyr/apStar-dr17-2M07534345+1916240.fits}{2M07534345+1916240} & SZ-Gem               & 118.431251 & 19.273111  & 10.898 & 1 & RRab        \\
\href{https://dr17.sdss.org/sas/dr17/apogee/spectro/redux/dr17/stars/apo1m/rrlyr/apStar-dr17-2M08045356+1945105.fits}{2M08045356+1945105} & IW-Cnc               & 121.223205 & 19.752972  & 11.567 & 1 & RRab        \\
\href{https://dr17.sdss.org/sas/dr17/apogee/spectro/redux/dr17/stars/apo1m/rrlyr/apStar-dr17-2M08062559+2315056.fits}{2M08062559+2315056} & SS-Cnc               & 121.606636 & 23.251583  & 11.146 & 1 & RRab        \\
\href{https://dr17.sdss.org/sas/dr17/apogee/spectro/redux/dr17/stars/apo1m/rrlyr/apStar-dr17-2M08133883+2110576.fits}{2M08133883+2110576} & CSS-J081338.8+211058 & 123.411743 & 21.182777  & 11.690  & 1 & RRc        \\
\href{https://dr17.sdss.org/sas/dr17/apogee/spectro/redux/dr17/stars/apo1m/rrlyr/apStar-dr17-2M08325518+1311285.fits}{2M08325518+1311285} & TT-Cnc               & 128.229553 & 13.191112  & 9.968  & 1 & RRab        \\
\href{https://dr17.sdss.org/sas/dr17/apogee/spectro/redux/dr17/stars/apo1m/rrlyr/apStar-dr17-2M08394723+1417243.fits}{2M08394723+1417243} & ASAS-J083947+1417.4  & 129.946793 & 14.290056  & 10.910  & 1 & RRc        \\
\href{https://dr17.sdss.org/sas/dr17/apogee/spectro/redux/dr17/stars/apo1m/rrlyr/apStar-dr17-2M09130504+4329160.fits}{2M09130504+4329160} & CSS-J091304.9+432915 & 138.270614 & 43.487694  & 10.459 & 1 & RRab        \\
\href{https://dr17.sdss.org/sas/dr17/apogee/spectro/redux/dr17/stars/apo1m/rrlyr/apStar-dr17-2M04171719+4724006.fits}{2M04171719+4724006} & AR-Per               & 64.321701 & 47.400200  & 8.680   & 5 & RRab        \\
\href{https://dr17.sdss.org/sas/dr17/apogee/spectro/redux/dr17/stars/apo1m/rrlyr/apStar-dr17-2M03083089+1026452.fits}{2M03083089+1026452} & X-Ari                & 47.128700 & 10.445900  & 7.944  & 7 & RRab        \\
\href{https://dr17.sdss.org/sas/dr17/apogee/spectro/redux/dr17/stars/apo1m/rrlyr/apStar-dr17-2M19252793+4247040.fits}{2M19252793+4247040} & RR-Lyr$^{\ast}$               & 291.365997 & 42.784401  & 6.693  & 7 & RRab        \\
\href{https://dr17.sdss.org/sas/dr17/apogee/spectro/redux/dr17/stars/apo1m/rrlyr/apStar-dr17-2M01320817+0120301.fits}{2M01320817+0120301} & RR-Cet               & 23.0340003 & 1.341750  & 8.688  & 8 & RRab        \\
\href{https://dr17.sdss.org/sas/dr17/apogee/spectro/redux/dr17/stars/apo1m/rrlyr/apStar-dr17-2M14163658+4221356.fits}{2M14163658+4221356} & TV-Boo               & 214.151993 & 42.359901  & 10.222 & 9 & RRc        \\
\href{https://dr17.sdss.org/sas/dr17/apogee/spectro/redux/dr17/stars/apo1m/rrlyr/apStar-dr17-2M00234308+2924036.fits}{2M00234308+2924036} & SW-And$^{\ast}$               & 5.929540 & 29.400999  & 8.517  & 11 & RRab       \\
\href{https://dr17.sdss.org/sas/dr17/apogee/spectro/redux/dr17/stars/apo1m/rrlyr/apStar-dr17-2M07272799+7242124.fits}{2M07272799+7242124} & EW-Cam               & 111.866996 & 72.703499  & 8.250   & 13 & RRab       \\
\href{https://dr17.sdss.org/sas/dr17/apogee/spectro/redux/dr17/stars/apo1m/rrlyr/apStar-dr17-2M09030779+4435082.fits}{2M09030779+4435082} & TT-Lyn$^{\ast}$               & 135.781997 & 44.585602  & 8.594  & 15 & RRab       \\
\href{https://dr17.sdss.org/sas/dr17/apogee/spectro/redux/dr17/stars/apo1m/rrlyr/apStar-dr17-2M11361176+8117369.fits}{2M11361176+8117369} & CN-Cam$^{\ast}$               & 174.048995 & 81.293602  & 8.353  & 15 & RRab       \\
\href{https://dr17.sdss.org/sas/dr17/apogee/spectro/redux/dr17/stars/apo1m/rrlyr/apStar-dr17-2M11294849+3004025.fits}{2M11294849+3004025} & TU-UMa$^{\ast}$               & 172.451995  & 30.067301  & 8.938  & 18 & RRab       \\
\href{https://dr17.sdss.org/sas/dr17/apogee/spectro/redux/dr17/stars/apo1m/rrlyr/apStar-dr17-2M09532839+0203263.fits}{2M09532839+0203263} & T-Sex$^{\ast}$                & 148.367996 & 2.057320  & 9.286  & 19 & RRc       \\
\href{https://dr17.sdss.org/sas/dr17/apogee/spectro/redux/dr17/stars/apo1m/rrlyr/apStar-dr17-2M22391317+6451305.fits}{2M22391317+6451305} & RZ-Cep$^{\ast}$               & 339.804993 & 64.858498  & 8.038  & 27 & RRc       \\
\href{https://dr17.sdss.org/sas/dr17/apogee/spectro/redux/dr17/stars/apo1m/rrlyr/apStar-dr17-2M22152563+0649214.fits}{2M22152563+0649214} & DH-Peg$^{\ast}$               & 333.856995 & 6.822630  & 8.643  & 49 & RRc     \\ \hline
\multicolumn{7}{l}{$\dagger$ Marked as RR~Lyrae variable in APOGEE DR17, but closer examination using the ASAS-SN photometry and \textit{Gaia} RR~Lyrae} \\
\multicolumn{7}{l}{catalog showed no luminosity variation that would justify classification as an RR~Lyrae pulsator.}
\end{tabular}
\end{table*}

\begin{table*}[]
\caption{Same as Table~\ref{tab:OtherRR} but for other nearby RR~Lyrae stars observed by the APOGEE survey used in the creation of the line-of-sight velocity template for RR~Lyrae stars.}
\label{tab:OtherRR}
\begin{tabular}{lcrrrrl}
\hline
APOGEE-ID & Alt. ID & RA & DEC & H & NVISITS & Type\\ \hline
\href{https://dr17.sdss.org/sas/dr17/apogee/spectro/redux/dr17/stars/apo1m/rrlyr/apStar-dr17-2M13033110+7106441.fits}{2M13033110+7106441} & OW-Dra$^{\ast}$                       & 195.879590  & 71.112251  & 9.317  & 27 & RRc     \\
\href{https://dr17.sdss.org/sas/dr17/apogee/spectro/redux/dr17/stars/apo1m/rrlyr/apStar-dr17-2M12083507-0027243.fits}{2M12083507-0027243} & UU-Vir$^{\ast}$                       & 182.146136 & -0.456759  & 9.503  & 8 & RRab       \\
\href{https://dr17.sdss.org/sas/dr17/apogee/spectro/redux/dr17/stars/apo1m/rrlyr/asStar-dr17-2M00283385-7210088.fits}{2M00283385-7210088} & CO-Tuc$^{\ast}$                       & 7.141047   & -72.169113 & 13.055 & 19 & RRc      \\
\href{https://dr17.sdss.org/sas/dr17/apogee/spectro/redux/dr17/stars/apo1m/rrlyr/asStar-dr17-2M00242224-7722083.fits}{2M00242224-7722083} & YZ-Hyi$^{\ast}$                       & 6.092692   & -77.368996 & 14.115 & 14 & RRab      \\
\href{https://dr17.sdss.org/sas/dr17/apogee/spectro/redux/dr17/stars/apo1m/rrlyr/apStar-dr17-2M12211673+0022029.fits}{2M12211673+0022029} & UV-Vir$^{\ast}$                       & 185.319722 & 0.367496 & 10.949 & 12 & RRab \\ \hline
\end{tabular}
\end{table*}

\begin{table}
\caption{Table with root-mean-square values along the pulsation phase (first column) for the template curves of RRab pulsators (second column) and RRc stars (third column) for data from the APOGEE survey.}
\label{tab:TEMPerr}
\begin{tabular}{lcc}
\hline
Phase & $\texttt{Temp}_{\rm err}^{\rm RRab}$ & $\texttt{Temp}_{\rm err}^{\rm RRc}$ \\ \hline
0.000 & 0.054 & 0.083 \\
0.025 & 0.054 & 0.083 \\
0.075 & 0.059 & 0.087 \\
0.125 & 0.064 & 0.115 \\
0.175 & 0.028 & 0.081 \\
0.225 & 0.064 & 0.068 \\
0.275 & 0.056 & 0.074 \\
0.325 & 0.040 & 0.069 \\
0.375 & 0.033 & 0.080 \\
0.425 & 0.044 & 0.108 \\
0.475 & 0.073 & 0.127 \\
0.525 & 0.046 & 0.101 \\
0.575 & 0.049 & 0.078 \\
0.625 & 0.069 & 0.117 \\
0.675 & 0.051 & 0.120 \\
0.725 & 0.066 & 0.092 \\
0.775 & 0.057 & 0.149 \\
0.825 & 0.050 & 0.391 \\
0.875 & 0.159 & 0.220 \\
0.925 & 0.186 & 0.215 \\
0.975 & 0.070 & 0.109 \\
1.000 & 0.070 & 0.109 \\ \hline
\end{tabular}
\end{table}

\begin{table*}
\caption{Table for $34$ RRab type variables with the lowest $\text{Amp}_{\rm los}$ located at the nonlinear part of the amplitude scaling relation. The first column contains the \textit{Gaia} identifier. The following three columns contain the photometric and line-of-sight velocity amplitudes and the pulsation periods of a given variable. The last column contains the photometric metallicities listed in the \textit{Gaia} RR Lyrae catalog \citep{Clementini2023}.}
\label{tab:SevenLow}
\begin{tabular}{lcccr}
\hline 
\texttt{source\_id} & $\text{Amp}_{G}$ & $\text{Amp}_{\rm los}$ & $P$ & [Fe/H]$_{\rm phot}$ \\
 & [mag] & [km\,s$^{-1}$] & [day] & [dex] \\ \hline
  3181149548774349824 & $0.395 \pm 0.004$ & $34.8 \pm 5.0 $ & $0.64004$ & $ -0.73 \pm     0.23 $ \\
  3478166191064353920 & $0.312 \pm 0.003$ & $41.1 \pm 5.6 $ & $0.64333$ & $ -0.21 \pm     0.24 $ \\
  4957668453282872448 & $0.179 \pm 0.002$ & $24.8 \pm 3.3 $ & $0.71747$ & $ 0.31  \pm     0.27 $ \\
  4793950312913699328 & $0.376 \pm 0.016$ & $43.0 \pm 10.0$ & $0.61026$ & $ -0.21 \pm     0.28 $ \\
  5222811627876331648 & $0.316 \pm 0.004$ & $48.8 \pm 8.9 $ & $0.61568$ & $ -0.47 \pm     0.28 $ \\
  1326205915830514176 & $0.137 \pm 0.002$ & $15.2 \pm 5.5 $ & $0.61565$ & -- \\
  6164530246002222208 & $0.394 \pm 0.008$ & $44.5 \pm 4.6 $ & $0.62509$ & $ -0.77 \pm     0.24 $ \\
  3959667827791428224 & $0.292 \pm 0.003$ & $42.0 \pm 9.9 $ & $0.62838$ & $ -0.30 \pm     0.22 $ \\
  2198888054289887232 & $0.177 \pm 0.002$ & $20.5 \pm 5.1 $ & $0.44949$ & $ 0.07  \pm     0.22 $ \\
  4387211137548084352 & $0.395 \pm 0.008$ & $43.2 \pm 5.7 $ & $0.73768$ & $ -0.38 \pm     0.25 $ \\
  4137503800563739136 & $0.234 \pm 0.006$ & $31.4 \pm 7.8 $ & $0.60553$ & -- \\
  1677592858356166528 & $0.262 \pm 0.006$ & $33.7 \pm 3.5 $ & $0.80726$ & -- \\
  1954728333254872576 & $0.221 \pm 0.002$ & $27.1 \pm 1.2 $ & $0.67400$ & $ 0.33  \pm     0.22 $ \\
  5810405553887250432 & $0.356 \pm 0.005$ & $37.0 \pm 3.4 $ & $0.84691$ & $ -0.07 \pm     0.25 $ \\
  6409095201484462208 & $0.149 \pm 0.002$ & $23.5 \pm 3.9 $ & $0.80212$ & -- \\
  4984655725669340544 & $0.361 \pm 0.005$ & $55.3 \pm 6.9 $ & $0.61560$ & $ -0.54 \pm     0.23 $ \\
  1978708686175347968 & $0.214 \pm 0.001$ & $31.6 \pm 2.2 $ & $0.64911$ & $ -0.16 \pm     0.22 $ \\
  4702297875480363648 & $0.331 \pm 0.008$ & $40.2 \pm 1.6 $ & $0.72796$ & $ -0.26 \pm     0.30 $ \\
  3504228735513593600 & $0.203 \pm 0.002$ & $25.0 \pm 8.0 $ & $0.65251$ & $ -0.39 \pm     0.28 $ \\
  5459760747346789888 & $0.168 \pm 0.003$ & $26.0 \pm 4.0 $ & $0.69682$ & $ 0.31 \pm      0.32 $ \\
  5996105366938238336 & $0.321 \pm 0.054$ & $44.9 \pm 11.1$ & $0.58204$ & -- \\
  6151860053822639104 & $0.361 \pm 0.007$ & $39.7 \pm 1.1 $ & $0.63780$ & $ -0.78 \pm     0.30 $ \\
  5993797084032636928 & $0.189 \pm 0.008$ & $17.7 \pm 0.4 $ & $0.41930$ & -- \\
  4439524212872324736 & $0.183 \pm 0.002$ & $25.1 \pm 3.5 $ & $0.64206$ & $ -0.37 \pm     0.26 $ \\
  4605376543969500160 & $0.333 \pm 0.018$ & $39.4 \pm 8.8 $ & $0.58604$ & $ -0.43 \pm     0.35 $ \\
  6248312516647094528 & $0.179 \pm 0.010$ & $22.5 \pm 2.4 $ & $0.58327$ & -- \\
  538832101445230720  & $0.194 \pm 0.003$ & $26.8 \pm 1.0 $ & $0.64960$ & $ -0.10 \pm     0.26 $ \\
  2869411785820130304 & $0.316 \pm 0.013$ & $49.4 \pm 8.1 $ & $0.69608$ & $ 0.26 \pm      0.30 $ \\
  6019142407624043136 & $0.356 \pm 0.006$ & $44.6 \pm 10.3$ & $0.47148$ & $ 0.04 \pm      0.22 $ \\
  6529376963198996352 & $0.326 \pm 0.006$ & $37.2 \pm 7.7 $ & $0.89489$ & $ 0.48 \pm      0.45 $ \\
  1546016672688675200 & $0.365 \pm 0.006$ & $41.2 \pm 0.6 $ & $0.59958$ & $ -0.08 \pm     0.22 $ \\
  1223679342759114752 & $0.365 \pm 0.006$ & $48.0 \pm 3.8 $ & $0.76900$ & $ -0.06 \pm     0.29 $ \\
  5639451149541847552 & $0.360 \pm 0.012$ & $47.2 \pm 8.9 $ & $0.73313$ & -- \\
  4136943255792306304 & $0.230 \pm 0.004$ & $29.8 \pm 2.0 $ & $0.59972$ & $ 0.09 \pm      0.23 $ \\ \hline
\end{tabular}
\end{table*}

\begin{table}
\caption{Same as Table~\ref{tab:TEMPerr} but for scatter along the line-of-sight velocity templates derived based on \textit{Gaia} calcium triplet velocities.}
\label{tab:TEMPerrGaia}
\begin{tabular}{lcccc}
\hline
Phase & $\texttt{Temp}_{\rm err}^{\rm RRab-1}$ & $\texttt{Temp}_{\rm err}^{\rm RRab-2}$ & $\texttt{Temp}_{\rm err}^{\rm RRab-3}$ & $\texttt{Temp}_{\rm err}^{\rm RRc}$    \\ \hline
0.000 & 0.068 & 0.073 & 0.053 & 0.095 \\
0.025 & 0.068 & 0.073 & 0.053 & 0.095 \\
0.075 & 0.036 & 0.071 & 0.048 & 0.079 \\
0.125 & 0.054 & 0.052 & 0.047 & 0.105 \\
0.175 & 0.051 & 0.049 & 0.058 & 0.062 \\
0.225 & 0.065 & 0.033 & 0.044 & 0.082 \\
0.275 & 0.053 & 0.051 & 0.043 & 0.069 \\
0.325 & 0.069 & 0.047 & 0.039 & 0.064 \\
0.375 & 0.057 & 0.051 & 0.063 & 0.096 \\
0.425 & 0.069 & 0.050 & 0.058 & 0.052 \\
0.475 & 0.053 & 0.048 & 0.032 & 0.086 \\
0.525 & 0.056 & 0.051 & 0.023 & 0.064 \\
0.575 & 0.060 & 0.069 & 0.031 & 0.056 \\
0.625 & 0.057 & 0.055 & 0.110 & 0.061 \\
0.675 & 0.072 & 0.049 & 0.066 & 0.086 \\
0.725 & 0.045 & 0.052 & 0.071 & 0.165 \\
0.775 & 0.058 & 0.068 & 0.039 & 0.121 \\
0.825 & 0.095 & 0.073 & 0.093 & 0.163 \\
0.875 & 0.112 & 0.273 & 0.215 & 0.210 \\
0.925 & 0.218 & 0.172 & 0.036 & 0.091 \\
0.975 & 0.064 & 0.094 & 0.038 & 0.091 \\
1.000 & 0.064 & 0.094 & 0.038 & 0.091 \\ \hline 
\end{tabular}
\end{table}

\begin{table}
\caption{List of RR~Lyrae stars used in the creation of line-of-sight velocity templates from \textit{Gaia} RVS spectra. The columns present the \textit{Gaia} ID, pulsation period, and RR~Lyrae pulsation type.}
\label{tab:FourTEMPGaiaUsedStars}
\begin{tabular}{lcc}
\hline
\texttt{source\_id} & $P$\,[day] & Type \\ \hline
1317846466364172800 & $0.33168$ & RRc \\
2642479663953557888 & $0.30626$ & RRc \\
3406613410300235904 & $0.23686$ & RRc \\
1956531880222667904 & $0.2537$ & RRc \\
5697506806599947392 & $0.26977$ & RRc \\
5815008831122635520 & $0.37404$ & RRc \\
837077516695165824 & $0.34936$ & RRc \\
6884361748289023488 & $0.27346$ & RRc \\
1063808840251264128 & $0.2985$ & RRc \\ \hline
& \hspace{-2cm} RRab $P < 0.55$\,day & \\ \hline
1760981190300823808 & $0.47262$ & RRab1 \\
5431789686933763456 & $0.51657$ & RRab1 \\
5360400630327427072 & $0.52743$ & RRab1 \\
4596935593202765184 & $0.39961$ & RRab1 \\
5397395623185535104 & $0.53103$ & RRab1 \\
4467433017738606080 & $0.45536$ & RRab1 \\
4224859720193721856 & $0.36179$ & RRab1 \\
2093443102473433728 & $0.52527$ & RRab1 \\
1286188056265485952 & $0.37734$ & RRab1 \\
4539434124372063744 & $0.41138$ & RRab1 \\
182142003881848832 & $0.45609$ & RRab1 \\
4985455998336183168 & $0.51091$ & RRab1 \\
1853751148171392256 & $0.41986$ & RRab1 \\
2254366868398077312 & $0.47649$ & RRab1 \\ \hline
& \hspace{-2cm} RRab $0.55 < P < 0.7$\,day & \\ \hline
5801111519533424384 & $0.57972$ & RRab2 \\
856816808430505856 & $0.62731$ & RRab2 \\
3143813565573130880 & $0.55051$ & RRab2 \\
5461994302138361728 & $0.57435$ & RRab2 \\
2131968508140833920 & $0.61323$ & RRab2 \\
2981136563934324224 & $0.58725$ & RRab2 \\
2022835523801236864 & $0.59413$ & RRab2 \\
5413725466808434048 & $0.60507$ & RRab2 \\
3486473757325180032 & $0.6503$ & RRab2 \\
2973463347160718976 & $0.58147$ & RRab2 \\
1533880980593444352 & $0.69779$ & RRab2 \\
1483653713185923072 & $0.55176$ & RRab2 \\
6151860053822639104 & $0.6378$ & RRab2 \\
4124594086640564352 & $0.60259$ & RRab2 \\
1546016672688675200 & $0.59958$ & RRab2 \\
2976126948438805760 & $0.56991$ & RRab2 \\ \hline
& \hspace{-2cm} RRab $P > 0.7$\,day & \\ \hline
1677592858356166528 & $0.80726$ & RRab3 \\
3587566361077304704 & $0.73284$ & RRab3 \\
4702297875480363648 & $0.72796$ & RRab3 \\
6788454544456587520 & $0.73188$ & RRab3 \\
6046836528519375232 & $0.70928$ & RRab3 \\ \hline
\end{tabular}
\end{table}

\end{appendix}

\end{document}